\documentclass{article}
\usepackage[table,dvipsnames]{xcolor}
\usepackage{microtype}
\usepackage{graphicx}
\usepackage{booktabs} 
\usepackage{hyperref}
\usepackage[accepted]{mlsys2026}

\usepackage{tikz}
\usepackage{color}
\usepackage{amsmath}
\usepackage{balance}
\usepackage{subcaption}
\usepackage{url}
\usepackage{pifont}
\usepackage{siunitx}
\usepackage[utf8]{inputenc}
\usepackage{enumitem}
\usepackage[printonlyused]{acronym}
\usepackage{multirow}
\usepackage{afterpage}
\usepackage{multicol}
\usepackage{tcolorbox}
\usepackage{minted}
\usepackage{listings}
\usepackage{seqsplit}
\usepackage[framemethod=TikZ]{mdframed}
\usepackage{xcolor}
\usepackage[all]{nowidow}

\DeclareTextFontCommand{\mytexttt}{\ttfamily\hyphenchar\font=45\relax\spaceskip=0.5\fontdimen2\font}

\mlsystitlerunning{Blueprint, Bootstrap, and Bridge: A Security Look at NVIDIA GPU Confidential Computing}

\begin{document}
\newcounter{TODO}
\newcommand{\TODO}[1]{\textcolor{blue}{\{TODO-\arabic{TODO}: #1\}}\addtocounter{TODO}{1}}

\newcommand{\zg}[1]{\textcolor{violet}{\{zgu: #1\}}}
\newcommand{\mvle}[1]{\textcolor{red}{\{mvle: #1\}}}
\newcommand{\sx}[1]{\textcolor{green}{\{Shixuan: #1\}}}
\newcommand{\js}[1]{\textcolor{purple}{\{Julian: #1\}}}
\newcommand{\sa}[1]{\textcolor{magenta}{\{Salman: #1\}}}
\newcommand{\rv}[1]{\textcolor{orange}{\{rvaldez: #1\}}}
\newcommand{\ZQ}[1]{\textcolor{blue}{\{ZQ: #1\}}}

\newcommand{\revision}[1]{\textcolor{blue}{#1}}

\newcommand{\nip}[1]{\vspace{1ex}\noindent\textbf{#1}}

\newcommand{\eg}{\emph{e.g., }}
\newcommand{\ie}{\emph{i.e., }}
\newcommand{\etal}{\emph{et al. }}
\newenvironment{zeroindent}
  {\par\setlength{\parindent}{0pt}}
  {\par}

\newcommand{\creg}{\textsuperscript{\textregistered}} 
\newcommand{\ignore}[1]{}

\setlength{\fboxsep}{1pt} 
\definecolor{light-gray}{gray}{0.99}
\newcommand{\code}[1]{\texttt{#1}}
\newcommand{\spec}[1]{\textit{#1}}

\newcommand{\speculation}[1]{\begin{mdframed}[
    hidealllines=false,    
    backgroundcolor=yellow!20, 
    innerleftmargin=5pt, 
    innerrightmargin=5pt,
    innertopmargin=5pt,
    innerbottommargin=5pt
]
#1
\end{mdframed}}

\newcommand{\question}[1]{\begin{tcolorbox}[colback=green!10,colframe=black!50,boxrule=0.2pt, top=1pt, bottom=1pt, left=1pt, right=1pt] 
\textit{#1}
\end{tcolorbox}}

\newsavebox{\shortpagebox}
\makeatletter
\newcommand{\shortpage}[1]
{\par
  \setbox\shortpagebox=\vbox{\strut #1\par}%
  \afterpage{\onecolumn
    \begin{multicols}{2}
    \unvbox\AP@partial
    \end{multicols}}%
  \unvbox\shortpagebox
\par}
\makeatother

\newcommand{\halfcircle}{
\begin{tikzpicture}[scale=0.1]
  \draw (0,0) circle (1);
  \begin{scope}
    \clip (0,0) circle (1);
    \fill[black] (-1,-1) rectangle (0,1);
  \end{scope}
\end{tikzpicture}
}

\newcommand*{\redcircle}[1]{%
  \tikz[baseline=(char.base)]{
    \node[circle,fill=red,inner sep=1pt] (char)
      {\textcolor{white}{\bfseries\tiny #1}};
  }%
}

\newcommand*{\blackcircle}[1]{%
  \tikz[baseline=(char.base)]{
    \node[circle,fill=black,inner sep=1pt] (char)
      {\textcolor{white}{\bfseries\tiny #1}};
  }%
}

\twocolumn[
\mlsystitle{Blueprint, Bootstrap, and Bridge: \\A Security Look at NVIDIA GPU Confidential Computing}

\begin{mlsysauthorlist}
\mlsysauthor{Zhongshu Gu}{ibm}
\mlsysauthor{Enriquillo Valdez}{ibm}
\mlsysauthor{Salman Ahmed}{ibm}
\mlsysauthor{Julian James Stephen}{ibm}
\mlsysauthor{Michael V. Le}{ibm}
\mlsysauthor{Hani Jamjoom}{ibm}
\mlsysauthor{Shixuan Zhao}{ohio}
\mlsysauthor{Zhiqiang Lin}{ohio}
\end{mlsysauthorlist}

\mlsysaffiliation{ibm}{IBM Research}
\mlsysaffiliation{ohio}{The Ohio State University}

\mlsyscorrespondingauthor{Zhongshu Gu}{zgu@us.ibm.com}

\mlsyskeywords{GPU Confidential Computing}

\vskip 0.3in

\begin{abstract}
NVIDIA \ac{GPU-CC} aims to provide secure execution for AI workloads. For end users, enabling \ac{GPU-CC} is seamless and requires no modifications to existing applications. However, this ease of adoption relies on a proprietary and highly complex system that is difficult to inspect, creating challenges for researchers seeking to understand its architecture and security landscape. In this work, we provide a security look at \ac{GPU-CC} by reconstructing a coherent view of the system. We first examine the system's \emph{blueprint}, focusing on the specialized architectural engines that support its security mechanisms. We then analyze the \emph{bootstrap} process, which coordinates hardware and software components to establish these protections. Finally, we conduct targeted experiments to assess whether, under the \ac{GPU-CC} threat model, data transfers along different paths remain protected across the \emph{bridge} between trusted CPU and GPU domains. We responsibly disclosed all security findings presented in this paper to the NVIDIA \ac{PSIRT}.
\end{abstract}
]
\printAffiliationsAndNotice{}
\acresetall
\section{Introduction}
Confidential computing enables secure outsourcing of sensitive computations to \acp{TEE} on public clouds, protecting the confidentiality and integrity of data while in use. Early efforts focused on \ac{CPU-CC}, which secures computations within the CPU package and encrypts system memory. As modern workloads increasingly rely on GPUs, driven by the growth of large language and vision models, NVIDIA introduced \ac{GPU-CC}~\cite{dhanuskodi2023creating} starting with the \spec{Hopper} architecture, integrating support into CUDA and GPU kernel drivers. \ac{GPU-CC} extends the trust boundary from CPUs to GPUs, creating a unified protected domain for AI pipelines. This advancement expands support for secure GPU-accelerated applications and enables collaborative learning scenarios~\cite{mo2022ppfl,guo2024trustworthy,quoc2021secfl,eichner2024confidential,gu2019reaching,cheng2024deta} where mutually distrusting parties can jointly execute AI workloads without exposing private data.

\ac{GPU-CC} enables users to deploy AI workloads with no modifications to application-level programs or data. Despite this seamless user experience, \ac{GPU-CC} introduces extensive changes across the system software, firmware, and hardware stacks to secure communication between the CPU and GPU. For researchers in the machine learning system community, a comprehensive understanding of \ac{GPU-CC}'s architecture and security landscape is essential. This knowledge allows researchers to critically assess whether the environment satisfies specific security requirements under the stated threat model. However, achieving this depth of insight remains challenging due to three primary obstacles:

\emph{1. Lack of Public Specifications.} NVIDIA provides high-level overviews~\cite{dhanuskodi2023creating,gpuccdeploy} of \ac{GPU-CC} for end users, but leaves many technical details undefined. For example, \ac{GPU-CC}-related engines, such as the \ac{FSP}, \ac{GSP}, and \ac{SEC2}, only appear in diagrams, but their precise functions, interactions, and roles are largely undocumented. While NVIDIA's patents~\cite{rogers2023confidential1,rogers2023confidential2,rogers2025implementing} provide more technical details than white papers, their claims are written in legalistic, abstract terms to maximize protection and cover a broad range of possible implementations, making them difficult to interpret.

\emph{2. Proprietary Ecosystem.} While NVIDIA has open-sourced certain components, such as \spec{open GPU kernel modules}~\cite{ogkm} and \spec{nvTrust}~\cite{nvtrustrepo}, most elements related to \ac{GPU-CC} remain closed, either embedded within GPUs or distributed only as binary executables.

\emph{3. System Complexity.} 
\ac{GPU-CC} introduces substantial modifications across system stacks. To preserve compatibility with existing implementations, \ac{GPU-CC} layers new data protection mechanisms over heterogeneous and legacy components. For example, CPU-GPU data transfers traverse multiple distinct data paths, each with its own transmission mechanisms, data formats, and granularities. This compatibility-driven design significantly increases implementation complexity and broadens the attack surface, making it harder to reason about the system's security guarantees.

In this paper, we aim to reconstruct a coherent view of the \ac{GPU-CC} system and conduct security analysis structured around three stages: \emph{Blueprint}, \emph{Bootstrap}, and \emph{Bridge}.

\nip{Blueprint (\S\ref{sec:engines}).}
We begin by instrumenting and intercepting inter-component interactions within the \ac{GPU-CC} system. By correlating these experimental observations with fragmented information synthesized from public sources~\cite{dhanuskodi2023creating,gpuccdeploy,rogers2023confidential1,rogers2023confidential2,rogers2025implementing}, we uncover a static blueprint of the system, identifying the architectural engines that form the foundation of \ac{GPU-CC}'s security design.

\nip{Bootstrap (\S\ref{sec:bootstrap}).}
Next, we analyze the bootstrap process, including secure boot, key generation, firewall establishment, and device attestation, which bring \ac{GPU-CC} to a trustworthy state. We examine the dependencies between hardware and software components and study how security mechanisms are orchestrated and enforced during this transition.

\nip{Bridge (\S\ref{sec:protection}).}
Finally, we evaluate the security of runtime data transfers across the bridge between the trusted CPU and GPU domains. These transfers traverse the \emph{untrusted} PCIe interface, making them vulnerable to observation or manipulation by adversaries controlling the host system. For each data path, we investigate the protection mechanisms in place, analyze potential attack surfaces, assess their security implications, and propose possible mitigations. All security findings presented in this paper have been responsibly disclosed to NVIDIA's \ac{PSIRT}.
\section{Background}
In this section, we provide a brief overview of the evolution of confidential computing from CPU to GPU, summarize earlier research efforts on enabling confidential computing on GPUs, and discuss the current status of NVIDIA's \ac{GPU-CC} technology.

\nip{\ac{CPU-CC} Technologies.} NVIDIA \ac{GPU-CC} cannot operate independently. It must work in conjunction with \ac{VM}-based \ac{CPU-CC}, such as Intel \ac{TDX}~\cite{tdxwhitepaper,cheng2024intel} and AMD \ac{SEV}~\cite{kaplan2016amd,kaplan2017protecting,sev2020strengthening}.  \ac{CPU-CC} protects the confidentiality and integrity of computations and data within \acp{CVM} by enforcing cryptographic isolation from the host environment, including the hypervisor, system administrators, and I/O devices, which are considered untrusted or potentially compromised. This protection relies on security features such as trusted \ac{VM} management and runtime memory encryption. Each \ac{CVM} is assigned a unique ephemeral key that is used to encrypt the memory. The \ac{AES} engine in the on-die memory controller handles the encryption and decryption of data transferred between the CPU and the \ac{CVM}'s private memory. \ac{CPU-CC} also supports remote attestation to verify the authenticity and integrity of the trusted platform. 

\nip{I/O Protection in \ac{CPU-CC}.} The trust boundary of \ac{CPU-CC} is limited to the CPU and its designated private memory regions. When data leaves this boundary, e.g., during I/O operations, it must pass through untrusted components such as the hypervisor and I/O devices. I/O devices commonly use \ac{DMA} to read and write memory without CPU involvement. Since these devices are untrusted, they are not permitted direct access to the \ac{CVM}'s private memory, which is encrypted with the \ac{CVM}'s key. The general practice is to employ software-based encryption mechanisms, e.g., \ac{TLS} is used for network traffic and \ac{LUKS} is used for disk encryption. In this model, I/O-bound data is first encrypted within the \ac{CVM}'s private memory, then transferred to a staging buffer. This buffer, which is outside the \ac{CVM}'s encrypted memory, is accessible via \ac{DMA} by I/O devices. Therefore, the encrypted I/O data can be securely transmitted to the devices. 

Unlike network interface cards or storage disks, GPUs are not I/O devices solely responsible for data delivery. Instead, they have evolved into general-purpose compute devices composed of numerous \acp{SM} and equipped with on-package \ac{HBM}. For the GPU to process offloaded code and data, these assets must ultimately be in plaintext within the GPU's device memory at runtime. Meanwhile, the PCIe interconnect between the CPU and GPU is generally considered untrusted and is vulnerable to snooping and tampering attacks. Therefore, extending the trust boundary of confidential computing from CPUs to GPUs requires establishing end-to-end secure channels that protect code and data as they traverse the untrusted interconnects.

\nip{Prior Research on \ac{GPU-CC}.} Research efforts to support \ac{GPU-CC} began with systems like Graviton~\cite{volos2018graviton} and HIX~\cite{jang2019heterogeneous}. Graviton proposed modifications to the GPU's command processor, while HIX introduced changes to the I/O interconnect. However, due to the proprietary nature of NVIDIA GPUs, Graviton's approach could only be evaluated via emulation, and HIX lacks protection against hardware-based attacks on PCIe and GPU memory. To avoid hardware modifications, researchers have also explored software-based \ac{GPU-CC} approaches. For instance, SAGE~\cite{ivanov2023sage} used Intel \ac{SGX} as a local verifier to establish a dynamic root of trust on the GPU, while Honeycomb~\cite{mai2023honeycomb} performed static analysis to validate GPU applications at load time. Confidential computing has also been explored for other accelerators such as IPUs~\cite{vaswani2023confidential} and NPUs~\cite{lee2022tnpu}, where there is generally more flexibility in customizing the hardware. In addition to x86 systems, \ac{GPU-CC} has also been studied in the context of Arm systems~\cite{wang2024cage,jiang2022cronus,deng2022strongbox}. More broadly, Wang and Oswald~\cite{wang2026confidential} provide a comprehensive survey of recent advances in confidential computing for CPU-GPU systems.

One of the greatest challenges in designing \ac{GPU-CC} stems from the immutability of commodity GPUs. While modifying hardware and firmware is the most direct path to enabling secure execution on GPUs, it is often impractical. However, GPU vendors like NVIDIA possess the flexibility to add confidential computing functionality directly into their hardware/firmware. NVIDIA introduced the first commercial \ac{GPU-CC} solution~\cite{dhanuskodi2023creating} as part of its \spec{Hopper} architecture. Recent research has started to focus on improving the performance of NVIDIA \ac{GPU-CC}~\cite{mohan2024securing,tan2025pipellm,tan2024performance, zhao2025gpu,lee2025characterization, wang2024fastrack}.
However, a significant gap remains between using the \ac{GPU-CC} system and understanding its inner workings, largely due to the closed nature of NVIDIA's ecosystem. This lack of transparency makes it difficult to evaluate whether NVIDIA \ac{GPU-CC} meets the security requirements of confidential computing or if its implementation can withstand various attack vectors. 

Our study focuses on \ac{GPU-CC} in NVIDIA H100 GPUs. Certain new \ac{GPU-CC} features, such as Multi-GPU support and Trusted I/O, are disabled or unavailable on our test platform. We discuss these features conceptually in Appendix~\S\ref{sec:future}. 
\section{Threat Model}
NVIDIA's \ac{GPU-CC} threat model \emph{inherits} and \emph{extends} the threat assumptions of \ac{VM}-based \ac{CPU-CC} technologies such as Intel \ac{TDX} and AMD \ac{SEV}.

\nip{Adversarial Capabilities under \ac{CPU-CC}.} Adversaries are assumed to have physical or remote access to the host machine, potentially controlling the boot firmware, \ac{SMM}, host operating system, hypervisor, and peripheral devices. They may also access host system memory.

\nip{Extended Adversarial Capabilities under \ac{GPU-CC}.} 
Adversaries may monitor and manipulate the traffic on the CPU-GPU interconnects. They can interact with GPU configurations via in-band tools (e.g., \spec{nvTrust}~\cite{nvtrustrepo}) or out-of-band interfaces like the \ac{BMC}. They may flash the GPU's VBIOS or update its firmware, reassign GPUs between \acp{CVM}, detach and reattach GPUs, or physically remove them from PCIe buses. 

\nip{Out of Scope.} \ac{CPU-CC} and \ac{GPU-CC} provide confidentiality and integrity guarantees for data in use, but do not address system availability. Adversaries with control over the host can launch denial of service attacks at will. Additionally, sophisticated physical attacks, such as decapsulation to probe internal interconnects or on-package \ac{HBM}, are also out of scope. Modern GPUs use advanced multi-layer packaging with dense interconnects and protective layers, making such attacks technically challenging and risky, with a high chance of permanently damaging the silicon.

\nip{Attack Surface.} 
As \ac{CPU-CC} and \ac{GPU-CC} are integrated into a unified security domain, the attack surface expands, allowing adversaries to launch targeted strikes from three primary perspectives:

\emph{1. Attacks on the \ac{VM}-based \ac{CPU-CC}.} In a typical deployment, a physical GPU is passed through to a \ac{CVM}, which hosts all sensitive code and data prior to GPU offloading. The \ac{CVM} also holds the full privileges required to configure the GPU, making it a critical component of the \ac{TCB} within the unified security domain. Consequently, any compromise of the \ac{CVM} effectively undermines the integrity of the \ac{GPU-CC}. Recent research on \ac{VM}-based \ac{CPU-CC} has uncovered a number of security flaws in both AMD \ac{SEV}-SNP~\cite{cohen2022amd, schluter2024wesee, schluter2024heckler, zhang2024cachewarp, schluter2025heracles, gast2025counterseveillance, schluter2025rmpocalypse, de2025badram} and Intel \ac{TDX}~\cite{aktas2023intel, wilke2024tdxdown, rauscher2025tdxploit, rauscher2026telescope}. These vulnerabilities span a wide range of attack vectors and require timely firmware and microcode updates to mitigate them within the \ac{CPU-CC}.

\emph{2. Attacks on the CPU-GPU Interconnects.} This scenario focuses on the confidentiality and integrity of data in transit between the \ac{CVM} and the GPU. In the threat model, interconnects such as PCIe are treated as untrusted, leaving them susceptible to traffic interception. Specifically, data paths involving \ac{RPC}, memory transfers, \ac{UVM} operations, memory scrubbing, fault delivery, and CUDA operations are at risk. While prior work demonstrated the feasibility of reconstructing deep learning models by snooping the memory and PCIe buses~\cite{hu2020deepsniffer, zhu2021hermes}, the advent of \ac{CPU-CC} and \ac{GPU-CC}, which encrypt traffic across the memory and PCIe buses respectively, makes such side-channel analysis more challenging.

\emph{3. Attacks on the GPU Hardware.} Beyond the interconnects, the GPU itself presents a significant attack surface. This survey~\cite{naghibijouybari2022microarchitectural} provided a comprehensive overview of how vulnerabilities manifest for systems integrating hardware accelerators prior to 2022. Recent studies have explored NVIDIA GPU-specific security flaws, including electromagnetic side channels~\cite{zhan2022graphics}, isolation issues in \ac{MIG}~\cite{zhang2023t}, cache-based side channels~\cite{zhang2024invalidate+}, and code injection attacks~\cite{guo2024gpu}. These exploits typically assume the adversary and victim share an execution domain with direct GPU access. Consequently, these attacks become particularly critical if an adversary can successfully breach the \ac{CVM}, as described in the first scenario.

\section{Methodology}
We analyze NVIDIA \ac{GPU-CC} through three stages, \emph{Blueprint} (\S\ref{sec:engines}), \emph{Bootstrap} (\S\ref{sec:bootstrap}), and \emph{Bridge} (\S\ref{sec:protection}), organized according to their chronological order and corresponding security properties. Each stage requires different methodologies for security analysis.

The \emph{Blueprint} stage corresponds to the static state of \ac{GPU-CC}, focusing on the functionalities and roles of the architectural engines. Since these engines operate behind the GPU boundary and their firmware is closed-source, their behavior cannot be directly monitored. NVIDIA patents~\cite{rogers2023confidential1,rogers2023confidential2,rogers2025implementing} provide more technical details than the white paper~\cite{dhanuskodi2023creating}, but the broad, generalized patent claims make them difficult to interpret. To narrow the scope, we infer engine functionalities by intercepting communications between NVIDIA software stacks and the hardware engines. For the NVIDIA kernel-mode and \ac{UVM} drivers, which are open-source, we instrument the code to track control flows to specific engines. For closed-source components, such as the CUDA runtime and user-mode driver, we preload modified libraries to indirectly observe execution events. We then cross-validate our findings against patent descriptions to refine the mapping of engine roles.

The \emph{Bootstrap} stage corresponds to the initialization of \ac{GPU-CC}, during which the system establishes a trustworthy state. We intercept the launch sequence managed by the NVIDIA kernel-mode driver to monitor how architectural engines are brought up. We also quantitatively measure the enforcement of security mechanisms (e.g., firewall) by evaluating the proportion of control registers that are blocked versus those still exposed. For device attestation, we extract attestation reports and certificate chains by intercepting related functions in \spec{nvTrust} and NVIDIA kernel-mode driver. The certificate hierarchy also helps validate the secure boot sequence.

The \emph{Bridge} stage corresponds to the runtime execution of \ac{GPU-CC}, focusing on the security of data transfers between trusted CPU and GPU domains. Because data flows traverse multiple distinct data paths with different implementations, we discuss the specific methodology case by case in \S\ref{sec:protection}.

\section{Platform Configuration}
\label{sec:platform}
Our host system is equipped with dual-socket AMD EPYC 9634 84-core processors, with \ac{SEV}-SNP enabled, and an eight-GPU NVIDIA H100 SXM5 setup. Each GPU has 80 GB of GPU memory, and the VBIOS version is \code{96.00.61.00.01}. The host operating system is Ubuntu \code{22.04.5} LTS running the Linux \code{5.19.0} kernel.

At the time of writing, \ac{GPU-CC} on our host does not support the Multi-GPU configuration. Therefore, we passed through a single GPU to a guest \ac{CVM} for testing. We evaluated two \acp{CVM}. One guest runs Red Hat Enterprise Linux 9.4 with the Linux \code{6.1.91} kernel, using NVIDIA driver version \code{550.54.15} and CUDA \code{12.4}. The other guest runs Ubuntu \code{22.04} with Linux kernel \code{6.8.0}, NVIDIA driver version \code{570.86.15}, and CUDA \code{12.8}. Each \ac{CVM} is provisioned with 64 GB of system memory.
\acresetall
\section{Blueprint: Architectural Engines}
\label{sec:engines}
\begin{figure}[!t]
\centering
\includegraphics[width=0.5\textwidth]{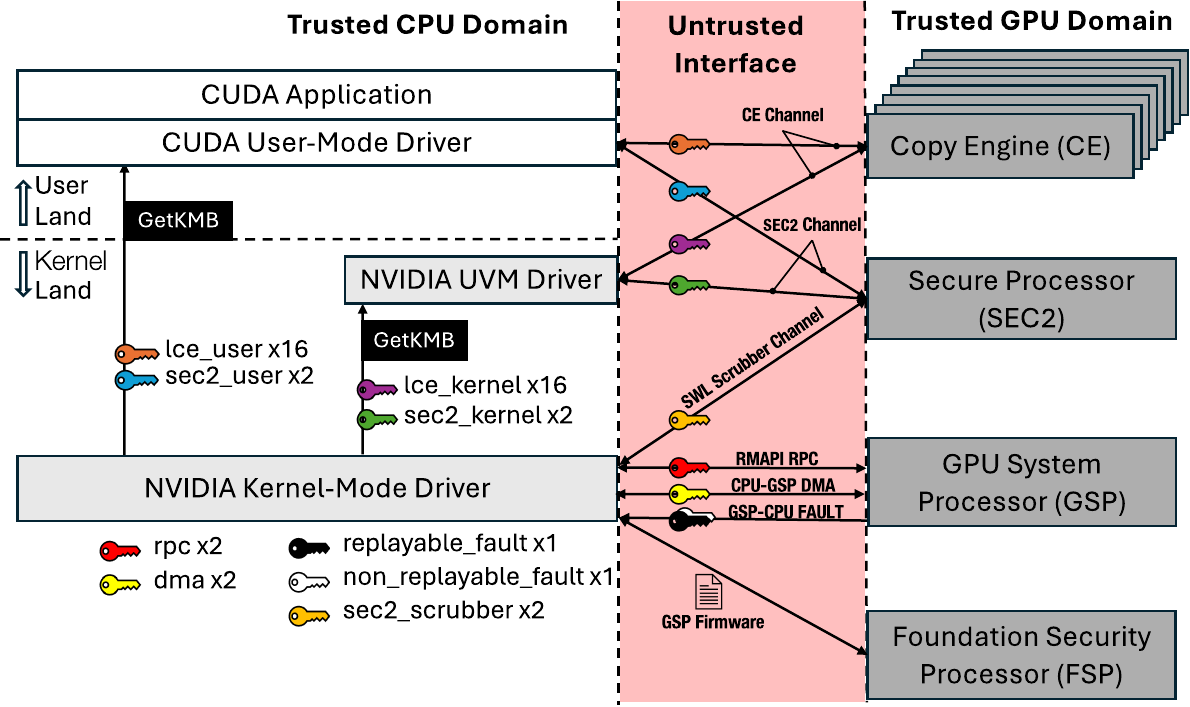}
\caption{The software/hardware components in \ac{GPU-CC}}
\label{fig:elements}
\end{figure}

Multiple architectural engines have been designed or repurposed to support \ac{GPU-CC}. As shown in Figure~\ref{fig:elements}, these engines interact directly with NVIDIA software stacks, such as the NVIDIA kernel-mode driver, \ac{UVM} driver, and CUDA user-mode driver. We outline the key functionalities of each engine and their roles in \ac{GPU-CC}.

\nip{\acf{FSP}.}
\ac{FSP} is characterized as a RISC-V microcontroller that undergoes a secure boot early in the hardware chain of trust. On the NVIDIA \spec{Hopper} architecture, once the \ac{FSP} is securely initialized, the NVIDIA kernel-mode driver transmits several binary images, specifically the \ac{GSP-FMC} and the \ac{GSP-RM}, across a dedicated communication channel. These images are digitally signed by NVIDIA to ensure integrity and authenticity. The \ac{FSP} validates the signatures before initializing and launching \ac{GSP-RM} on the \ac{GSP}.

In addition, in the GTC talk~\cite{gtcspring}, \ac{FSP} was described as responsible for setting up \ac{GPU-CC} mode and performing attestation. However, NVIDIA patents~\cite{rogers2023confidential1,rogers2023confidential2,rogers2025implementing} attribute these same operations to the \ac{SEC2} engine. Because the GPU-side implementation remains proprietary, the exact technical details are unclear. Nevertheless, since both the \ac{FSP} and \ac{SEC2} reside within the trusted boundary, from a security standpoint, it makes no difference which engine performs
these tasks. The shift likely reflects a migration of functionality during NVIDIA's architectural evolution.

\nip{\acf{GSP}.}
The \ac{GSP} serves as the GPU's control plane, handling initialization and resource management tasks. The \ac{GSP} engine is a RISC-V microcontroller equipped with \ac{AES} hardware for encryption and decryption. It communicates exclusively with the NVIDIA kernel-mode driver. In the context of \ac{GPU-CC}, the \ac{GSP} serves two main roles:

\emph{1. Key Negotiation.} When \ac{GPU-CC} is enabled, a \ac{SPDM} session is established before launching the \ac{GSP-RM}. The \ac{GSP} hosts an \spec{\ac{SPDM} Responder}, and during the handshake process, the NVIDIA kernel-mode driver negotiates a shared master secret with the \spec{\ac{SPDM} Responder}. This secret serves as the foundation for deriving all session keys that protect the data transmitted between the \ac{CVM} and GPU. 

\emph{2. \ac{RMAPI} \ac{RPC}.} Once the \ac{GSP-RM} is launched, it communicates with the NVIDIA kernel-mode driver via the \ac{RMAPI} \ac{RPC} channel and transfers memory through the CPU-\ac{GSP} \ac{DMA} channel. Separate session keys, derived from the master secret, protect data transmitted through the bi-directional \ac{RPC} and \ac{DMA} channels. Additionally, firmware running within the \ac{GSP} may sanitize all inputs and outputs to prevent unauthorized access or tampering by the host. 

\nip{\acf{SEC2}.}
\ac{SEC2} is a RISC-V microcontroller specifically designed for \ac{GPU-CC}. \ac{SEC2} can decrypt and verify data, but it cannot encrypt data. Unlike \ac{GSP}, \ac{SEC2} is accessible to both kernel-mode clients (e.g., NVIDIA kernel-mode driver and \ac{UVM} driver) and user-mode clients (e.g., CUDA user-mode driver). Based on the function descriptions in NVIDIA patents~\cite{rogers2023confidential1,rogers2023confidential2,rogers2025implementing}, \ac{SEC2} provides multiple security functionalities:

\emph{1. Setup of \ac{CPR} and \ac{GPU-CC} Mode.}
Evidence indicates that \ac{SEC2} is responsible for creating the \ac{CPR} and transitioning the GPU into \ac{GPU-CC} mode. The \ac{GPU-CC} mode setting can be stored in an EEPROM attached to the GPU but takes effect only after the next GPU reset. 

\emph{2. Device Attestation.}
Evidence indicates that \ac{SEC2} is responsible for generating an attestation report to prove the authenticity and trustworthiness of the GPU. The \spec{\ac{DIK}} is burned into the key fuse during silicon manufacturing, which is unique and immutable. The \spec{\ac{DIK}} is exclusively accessible by \ac{SEC2}, preventing access from any other entities. The attestation report is signed using an \ac{AK} derived from a hierarchy of cryptographic identities embedded in the silicon. 

\emph{3. Secure Data Transmission.}
\ac{SEC2} engine has session keys with the \ac{CVM} to protect the data in transit. The \ac{CVM} may either sign or encrypt the data using the session keys before placing it in an unprotected staging buffer. The \ac{SEC2} engine then retrieves the data: If it is signed, \ac{SEC2} verifies the integrity to detect any tampering during transit. If it is encrypted, \ac{SEC2} decrypts the data and stores it in the GPU's \ac{CPR}. 

\emph{4. Memory Scrubbing.}
Sensitive data within the GPU's \ac{CPR} must be erased when no longer needed. The NVIDIA kernel-mode driver can establish a \emph{scrubber channel} associated with the \ac{SEC2} engine. When physical memory pages are freed and require scrubbing, commands submitted to the scrubber channel are signed to prevent tampering. \ac{SEC2} can trigger a soft reset of the GPU, ensuring that before the GPU memory becomes visible on the system bus, all session keys are deleted, and memory is wiped. 

\emph{5. Secure Workload Submission.}
\ac{SEC2} can be used for bootstrapping secure workload submission. The CUDA user-mode driver and \ac{UVM} driver can allocate secure channels bound to the \ac{SEC2} engine, ensuring command integrity. Commands in this channel can be signed using an HMAC session key. \ac{SEC2} verifies the integrity of these commands before execution. 

\nip{\Acf{CE}.}
\acp{CE} manage memory transfers for GPUs and include \ac{AES} hardware for encryption and decryption. A GPU has one or more logical \acp{CE} and physical \acp{CE}. Physical \acp{CE} handle data movement, while logical \acp{CE} manage the control logic for physical \acp{CE}.
Logical \acp{CE} are accessible to both kernel-mode clients (e.g., NVIDIA kernel-mode driver and \ac{UVM} driver) and user-mode clients (e.g., CUDA user-mode driver). Clients can allocate secure channels to schedule work with logical \acp{CE}, with encryption keys obtained from the NVIDIA kernel-mode driver. Each logical \ac{CE} negotiates four keys with the kernel-mode driver in \ac{CVM}: two keys are for user mode and two for kernel mode. For each mode, one key protects host-to-device (\code{h2d}) data transfers, while the other protects device-to-host (\code{d2h}) transfers. All channels bound to the same logical \ac{CE} share the same key. Here are the security functionalities of \acp{CE}:

\emph{1. Data Movement Protection.} \ac{CE} ensures that data copied from \ac{CPR} to non-\ac{CPR} memory is encrypted, while data transferred from non-\ac{CPR} to \ac{CPR} is decrypted and integrity-checked. For example, it can retrieve encrypted GPU pushbuffers or CUDA kernels from unprotected memory, decrypting them into \ac{CPR} before executing them. Conversely, \acp{CE} can encrypt and sign data in \ac{CPR} before moving them to unprotected memory, such as when the GPU sends encrypted and signed synchronization signals (e.g., tracking semaphores) to notify command completion. Clients can then decrypt and verify these signals to proceed. Additionally, they may also support plaintext transfers of \ac{CPR}-to-\ac{CPR} or non-\ac{CPR}-to-non-\ac{CPR} memory. 

\emph{2. Memory Access Control.}
\acp{CE}, in conjunction with the GPU's Memory Management Unit, enforce access control by restricting compute engines to \ac{CPR} memory. Once a compute engine accesses \ac{CPR}, it is prevented from accessing any other unprotected memory regions. Any attempt to access memory outside \ac{CPR} is blocked, triggering a memory fault. 

\emph{3. Replay Attack Prevention.}
AES's \acp{IV} are used to prevent replay attacks. While channels on the same \ac{CE} share the same \code{h2d} and \code{d2h} keys, each channel independently maintains its own pair of \acp{IV} for \code{h2d} and \code{d2h} and increments them after each encryption or decryption. As a result, attempts to replay the old ciphertext would fail due to authentication tag mismatches. 
\begin{table*}[ht!]
\centering
\caption{The complete list of derived keys in \ac{GPU-CC}}
\begin{tabular}{llll}
\hline
Key Name                                    & Engine                & Protected Communication                           & Number \\ \hline
\code{\{gsp\_cpu,cpu\_gsp\}\_locked\_rpc}         & \multirow{4}{*}{\ac{GSP}}  & RPCs between CPU and \ac{GSP}                    & 2x1    \\
\code{\{gsp\_cpu,cpu\_gsp\}\_dma}                 &                       & Memory transfers between CPU and \ac{GSP}                    & 2x1    \\
\code{gsp\_cpu\_replayable\_fault}                 &                       & Replayable faults sent from \ac{GSP} to CPU     & 1      \\
\code{gsp\_cpu\_non\_replayable\_fault}            &                       & Non-replayable faults sent from \ac{GSP} to CPU & 1      \\ \hline
\code{cpu\_sec2\_\{data,hmac\}\_\{user,kernel\}} & \multirow{2}{*}{\ac{SEC2}} & Channels between CPU and \ac{SEC2}               & 2x2    \\
\code{cpu\_sec2\_\{data,hmac\}\_scrubber}         &                       & Scrubber channels between CPU and \ac{SEC2}      & 2x1    \\ \hline
\code{lce\{x\}\_\{h2d,d2h\}\_\{user,kernel\}}     & \ac{CE}                    & Channels between CPU and \ac{CE}                 & 8x2x2  \\ \hline
\end{tabular}
\label{tab:keys}
\end{table*}
\section{Bootstrap: Security Initialization}
\label{sec:bootstrap}
This section explains the \ac{GPU-CC} initialization process, covering:
(1) secure boot of architectural engines required for \ac{GPU-CC}, (2) key negotiation and derivation, 
(3) firewall establishment to block regular access to GPU registers and \ac{CPR} via the \ac{BAR}, and 
(4) device attestation to establish the trustworthiness of \ac{GPU-CC}.

\subsection{Secure Boot of Architectural Engines}
\label{ssec:secureboot}
To protect against malicious firmware updates or VBIOS flashing, \ac{GPU-CC} enforces a secure boot process for architectural engines. All firmware components must be authenticated and cryptographically signed by a trusted authority. 

According to the information shared by an NVIDIA engineer on the developer forum~\cite{nvidiadevforum}, the chain-of-trust follows this sequence: CEC EROT (if present) $\rightarrow$ \ac{FSP} $\rightarrow$ \ac{GSP} $\rightarrow$ \ac{SEC2}. Here, CEC EROT refers to Microchip’s \spec{CEC1736}~\cite{cec1736}, a real-time controller, where EROT stands for ``External Root of Trust.'' The \spec{CEC1736} plays a critical role as a hardware Root of Trust, anchoring the secure boot process and establishing the system's chain-of-trust. Both CEC EROT and \ac{FSP}'s \ac{BROM} authenticate \ac{FSP} firmware. This dual-layer verification ensures both external and internal integrity checks before \ac{FSP} is fully booted.

\nip{\ac{GSP} Initialization.} The initialization of the \ac{GSP} is a multi-step, cryptographically secured sequence. This process is orchestrated by the NVIDIA kernel-mode driver and anchored from the \ac{FSP}. Once the \ac{FSP} has completed its own secure boot, the driver issues a command to \ac{FSP} to trigger the \ac{GSP} boot process. The chronological flow is as follows:

\emph{1. Image Extraction and Memory Allocation.} The kernel-mode driver extracts the \ac{GSP-FMC} and \ac{GSP-RM} binary images from their respective archives. The driver then allocates unprotected system memory to load these images. These firmware components are typically bundled with NVIDIA kernel-mode driver releases, facilitating flexible runtime updates to the \ac{GSP} firmware. During this phase, the driver also calculates the total secure internal memory required by the \ac{GSP} firmware and populates a metadata structure to define the necessary memory region sizes.

\emph{2. \ac{CoT} Payload Construction.} The driver builds a \ac{CoT} message payload containing the system memory address of the \ac{GSP-FMC} image along with its authentication fields (cryptographic hashes, public keys, and signatures). This payload is transmitted to the \ac{FSP} via \ac{MCTP}.

\emph{3. \ac{FSP} Authentication and Initialization.} The \ac{FSP} processes the \ac{CoT} payload, copies the \ac{GSP-FMC} image, and verifies its digital signature. Upon successful verification, the \ac{GSP-FMC} can begin to execute.

\emph{4. \ac{GSP-RM} Loading.} The \ac{GSP-FMC} reads the sizing metadata (generated in step 1) to allocate the necessary device memory in the GPU. It then copies the \ac{GSP-RM} image into the allocated memory region and cryptographically validates its authenticity and integrity.

\emph{5. \ac{GSP} Engine Launch.} The \ac{GSP-FMC} points the hardware instruction pointer to the verified \ac{GSP-RM}'s entry point, effectively booting the \ac{GSP} engine. Once active, the \ac{GSP} initiates the key negotiation for \ac{GPU-CC} and establishes the protected \ac{RPC} channel for CPU-\ac{GSP} communication.

\nip{The Relationship between \ac{GSP} and \ac{SEC2}.} Although NVIDIA's forum~\cite{nvidiadevforum} indicates that the \ac{GSP} is responsible for starting the \ac{SEC2} engine, the precise boot sequence for these components remains opaque, as the internal handoff is concealed within the GPU hardware. However, analysis of the NVIDIA kernel-mode driver suggests that, starting with the \spec{Blackwell} architecture, the \ac{GSP} can also be initialized via the \ac{SEC2} engine using a process analogous to the \ac{FSP}-driven method. This shift likely represents a migration of security functionality, consolidating early-stage trust management within the \ac{SEC2} engine as part of NVIDIA's ongoing architectural iteration.

\subsection{Key Negotiation and Derivation}
\label{ssec:spdm}
The \spec{\acf{SPDM} Requester} within the NVIDIA kernel-mode driver initializes an \ac{SPDM} session with the \spec{\ac{SPDM} Responder} in the \ac{GSP}. This session establishes a shared master secret, from which all cryptographic keys used in \ac{GPU-CC} are derived to protect encrypted communications across the untrusted interface. 

Table~\ref{tab:keys} lists all the derived keys used in \ac{GPU-CC}, categorized by the engines they are associated with. For each key, we indicate the protected communication channel and the number of keys involved. The key names are designed to be self-explanatory, incorporating the engine name, direction, and key type. 

For \ac{GSP}, the following six keys are derived: the \code{gsp\_cpu\_locked\_rpc} and \code{cpu\_gsp\_locked\_rpc} keys are used to protect \ac{RMAPI} \ac{RPC} communication. The \code{gsp\_cpu\_dma} and \code{cpu\_gsp\_dma} keys are used to protect memory transfers between CPU and \ac{GSP}. Two additional keys, i.e., \code{gsp\_cpu\_replayable\_fault} and \code{gsp\_cpu\_non\_replayable\_fault}, protect the transmission of replayable and non-replayable fault packets from \ac{GSP} to CPU.

For \ac{SEC2}, six keys are derived to protect secure workload launch channels and scrubber channels. Data keys are used to encrypt data sent from the CPU to \ac{SEC2}. HMAC keys are used to sign data for integrity verification.
Each key is used in either user mode (e.g., by the CUDA user-mode driver) or kernel mode (e.g., by the \ac{UVM} driver). The
\code{cpu\_sec2\_data\_scrubber} and \code{cpu\_sec2\_hmac\_scrubber} keys are used by the scrubber channel. 

On an H100 GPU, secure data transfers are supported with eight logical \acp{CE}. Each engine utilizes distinct cryptographic keys to secure host-to-device and device-to-host transfers across both user and kernel modes. This architecture results in a total of 32 derived keys.

\subsection{Firewall Establishment}
\label{ssec:firewall}
\begin{figure}[!t]
\centering
\includegraphics[width=0.4\textwidth]{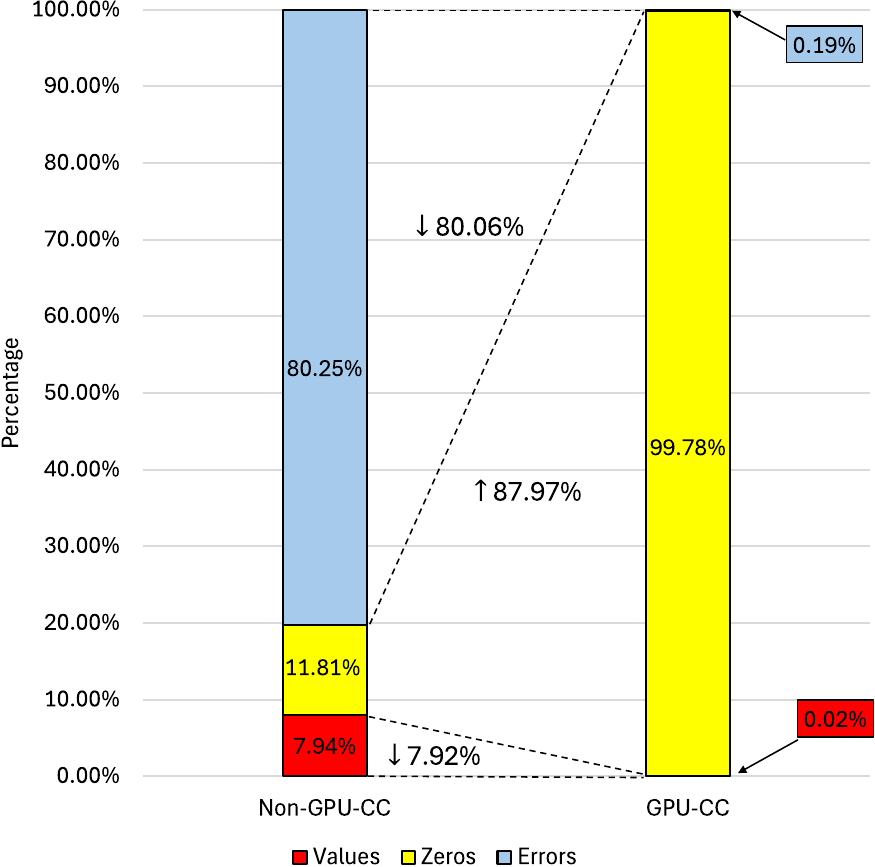}
\caption{Classification of register reads via GPU's BAR0}
\label{fig:bar0}
\end{figure}
In non-GPU-CC mode, system administrators on the host can freely access and modify the GPU's control and status registers, as well as its device memory, through \acp{BAR}. These registers configure various GPU settings, while the device memory may contain sensitive user data provisioned within the \ac{CVM}. Since the host, along with its administrators and software stacks, is no longer considered trusted in the threat model, \ac{BAR}-based PCIe access substantially enlarges the attack surface. 

To mitigate this, NVIDIA’s paper~\cite{dhanuskodi2023creating} states that enabling
\ac{GPU-CC} mode activates a PCIe firewall mechanism called \emph{BAR0 Decoupler}, which
blocks unauthorized accesses to the majority of registers through BAR0. However, a limited subset of registers remains accessible to support essential GPU management operations.

A critical concern is determining which registers remain accessible after \ac{GPU-CC} is enabled and whether their exposure poses any potential security risks. Unfortunately, the specifications for NVIDIA GPU control registers are proprietary, and no public documentation is available detailing the functionality of each register. Only a limited number of register fields have been referenced through the \spec{open GPU kernel modules} and \spec{nvTrust}.

Although the precise semantics of individual registers remain unclear, we quantitatively assess the difference in register accessibility before and after enabling the \ac{GPU-CC} mode. To this end, we developed a scanning program that simulates an adversary probing the entire BAR0 space on a H100 GPU under two modes, i.e., the non-GPU-CC mode and the \ac{GPU-CC} mode. The program reads from BAR0 starting at offset \code{0x0} and advances in 4-byte strides. Given that BAR0 spans 16 MB on a H100 GPU, this results in a total of \code{0x400000} read operations.

We categorize the returned values into three types: (1) \spec{Values}: non-zero numerical returns, (2) \spec{Zeros}: reads that return \code{0x0}, and (3) \spec{Errors}: reads returning error codes in the form of \code{0xbadxxxxx}. As shown in Figure~\ref{fig:bar0}, in non-GPU-CC mode, 7.94\% of the fields return values and 80.25\% return errors. In contrast, in \ac{GPU-CC} mode, 99.78\% of the fields return zero, with only 0.02\% (1,042 fields) returning non-zero values and 0.19\% returning errors. This significant zeroing effect is attributed to the activation of the \spec{BAR0 Decoupler}, which hides most control registers.

From an adversarial perspective, the decoupler obfuscates register visibility by returning zeros, making it difficult to determine whether a field is simply unmapped or protected. However, the small fraction (0.02\%) of accessible fields still raises security concerns. We advocate for transparency from NVIDIA regarding the functionality of these exposed registers and their necessity for GPU management.

Beyond register access, the PCIe firewall also blocks host access to the \acf{CPR} of GPU memory. This restriction leads to a rerouting of runtime data transmission between \acp{CVM} and GPUs, as discussed in detail in \S\ref{sec:protection}.

\subsection{Device Attestation}
\label{ssec:attestation}  


\begin{figure}[t!]
\centering
\includegraphics[width=0.5\textwidth]{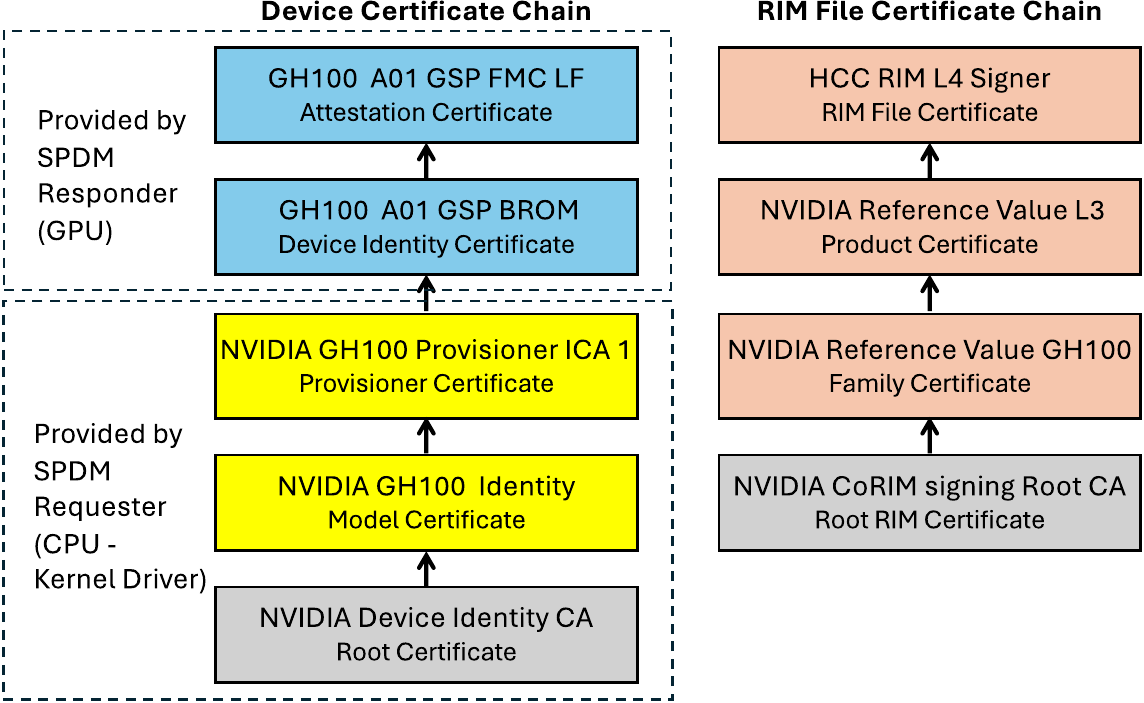}
\caption{GPU's device and RIM file certificate chains}
\label{fig:certificate_chain}
\end{figure}

The purpose of device attestation is to establish cryptographic proof that a \ac{CVM} is communicating with a genuine, uncompromised GPU.

\nip{Device Certificate Chain.} The process
begins with each NVIDIA GPU being provisioned with a unique
cryptographic identity: a hardware-fused \acf{DIK} and a
corresponding \spec{Device Identity Certificate}. This certificate, anchored to NVIDIA's Root \ac{CA}, forms the basis of a
certificate chain used to establish trust.

When a verifier running
inside a \ac{CVM} initiates attestation, it first retrieves the device certificate
chain. As illustrated in the left part of Figure~\ref{fig:certificate_chain}, the chain comprises
five certificates: the \spec{Attestation Certificate} and \spec{Device Identity Certificate} are retrieved from the GPU, while the remaining \spec{Provisioner
Certificate}, \spec{Model Certificate}, and \spec{Root Certificate} are obtained from
the NVIDIA kernel-mode driver.
The \spec{Device Identity Certificate} corresponds to the immutable \ac{GSP-BROM} and is signed by the NVIDIA Provisioner \ac{CA}.
The \spec{Device Identity Certificate} uniquely identifies the device as an authentic NVIDIA GPU. The immutable \ac{GSP-BROM} starts execution
and ensures that the system begins in a secure state. The \ac{GSP-FMC}
is loaded and authenticated. The GPU mathematically mixes its fused secret with specific firmware measurements to derive the \acf{AK} pair. The private \ac{AK} is used to sign attestation reports. The \spec{Attestation Certificate} is signed by the private \ac{DIK}, establishing a chain-of-trust. It contains the public \ac{AK} required to verify the attestation
report. The root of the certificate chain is the \spec{Root Certificate}, a
self-signed certificate that serves as the trust anchor.

The verifier replaces the \spec{Root Certificate} with a local one to prevent a compromised driver from undermining the certificate chain. It then verifies the chain in reverse order, starting with the \spec{Device Identity Certificate} and proceeding up to the \spec{Root Certificate}. During this process, the verifier sends multiple requests to the NVIDIA \ac{OCSP} service~\cite{ocspservice} to check the revocation status of each certificate. If the chain passes both the signature and revocation checks, the verifier proceeds to request the attestation report.

\nip{Attestation Report.} Once the certificate chain is validated, the verifier retrieves the attestation report directly from the GPU. This report is cryptographically signed by the private \ac{AK}. The verifier validates the attestation report's signature using the \spec{Attestation Certificate}. An attestation report contains measurement data and an opaque metadata block that identifies the GPU's firmware and driver. The measurement data consists of 64 structured records, each containing a measurement specification, a size field, and a cryptographic hash of a measured component (e.g., firmware, configuration). The opaque data block provides information to identify the GPU driver and firmware version. 

\nip{Measurements.} The verifier extracts the driver and VBIOS IDs from the opaque block and retrieves the corresponding \ac{RIM} files from the NVIDIA \ac{RIM} service~\cite{rimservice}. The verifier checks the status of the \ac{RIM} files by validating their schemas and endorsements. Each \ac{RIM} file includes a certificate chain. The verifier appends a local \spec{Root \ac{RIM} Certificate} to complete the chain (the right part of Figure~\ref{fig:certificate_chain}) and verify it. Once the chain is validated, the leaf certificate, i.e., \spec{\ac{RIM} File Certificate}, is used to verify the signature on the \ac{RIM} file. To validate the measurements, the verifier compiles a list of golden measurements from the \ac{RIM} files. It compares each measurement in the attestation report with the allowed values from the golden list. If a mismatch is found, the verifier concludes that the GPU is not in the expected state. If all checks succeed, the verifier reports the attestation result and the user can choose to transition the \ac{GPU-CC} into the \emph{READY} state.

\begin{table*}[!ht]
\centering
\caption{Classification of security implications by data paths}
\begin{tabular}{l|l|cc|l}
\hline
\textbf{Data Paths}               & \textbf{Engines}  & $\mathcal{C}$ & $\mathcal{I}$ & \textbf{Security Findings}                                                             \\ \hline
CPU-GSP \ac{RPC}              & \ac{GSP}      & \scalebox{1.5}{$\bullet$}   & \halfcircle  & Leakage and possible manipulations of \ac{RMAPI} \ac{RPC} invocations                                  \\ \hline
CPU-GSP Memory Transfers & \ac{GSP}      & \halfcircle  & \scalebox{1.5}{$\circ$}  & Timing channels for leaking memory transfer sizes                                 \\ \hline
GPU Memory Faults        & \ac{GSP}      & \halfcircle  &  \scalebox{1.5}{$\circ$} & Leakage of \code{PUT} pointer of shadow buffers                                                              \\ \hline
\ac{UVM}                      & \ac{SEC2}, \ac{CE} & \scalebox{1.5}{$\bullet$}  & \halfcircle  & Unprotected \code{GPFIFO}, \code{GPPUT}, and semaphores in \ac{SEC2} Channel                         \\ \hline
Memory Scrubbing         & \ac{SEC2}     & \scalebox{1.5}{$\bullet$}  & \halfcircle  & Unprotected semaphores in \ac{SEC2} Channel \\ \hline
CUDA                     & \ac{SEC2}, \ac{CE} & \textbf{?}  &  \textbf{?} & Closed-source CUDA runtime and user-mode driver                                   \\ \hline
\end{tabular}
\label{tab:category}
\end{table*}
\section{Bridge: Security Analysis of Data Transmission under \ac{GPU-CC}}
\label{sec:protection}
\begin{figure}[!t]
\centering
\includegraphics[width=0.5\textwidth]{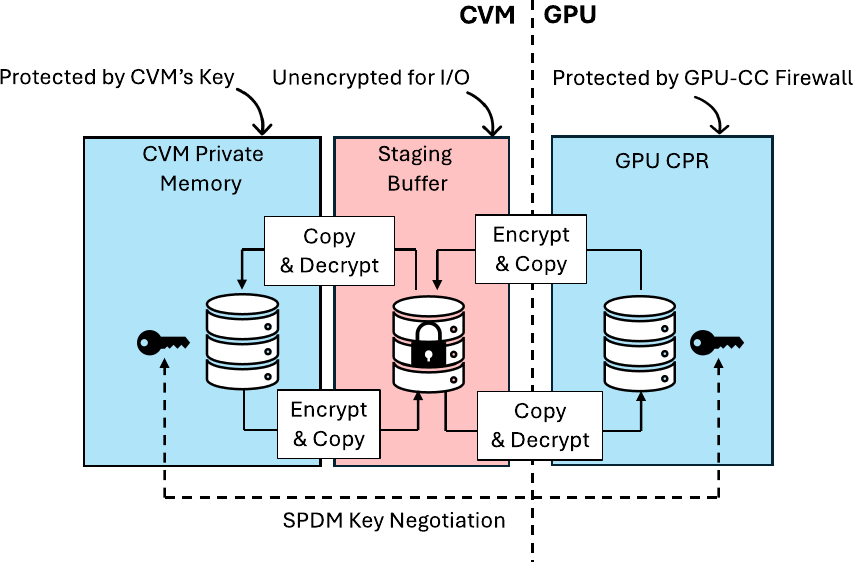}
\caption{Data transmission between \ac{CVM} and GPU}
\label{fig:memory}
\end{figure}

At runtime, data must be transmitted between the \ac{CVM}'s private memory and the GPU’s CPR over an untrusted PCIe interface, which can be intercepted or manipulated by adversaries on the host. This raises a key security question: 
\question{Does the data traversing this untrusted interface remain protected in terms of confidentiality and integrity?}
Data protection relies on the keys negotiated via \ac{SPDM} between the NVIDIA kernel-mode driver and the GPU. Different data paths employ different keys, with the full set of derived keys listed in \S\ref{ssec:spdm}. As shown in Figure~\ref{fig:memory}, staging buffers serve as intermediaries for these transfers. The staging buffer is a shared, unencrypted memory region accessible by both the \ac{CVM} and the GPU. Any data moving between protected regions, such as the \ac{CVM}'s private memory and the GPU's \ac{CPR}, must first be signed and/or encrypted using the negotiated key before being placed in the staging buffer. Upon retrieval, the data is verified and/or decrypted using the corresponding key.

By combining code analysis of NVIDIA kernel drivers with runtime monitoring of data transmission, we identified six types of data paths used to transfer data between the CVM and the GPU. Table~\ref{tab:category} presents these data paths along with their associated hardware engines. We further evaluated the security of each path in terms of confidentiality ($\mathcal{C}$) and integrity ($\mathcal{I}$). In the table, \scalebox{1.5}{$\bullet$} indicates a confirmed leakage or integrity violation, \scalebox{1.5}{$\circ$} denotes that no issues were observed, \halfcircle represents a potential vulnerability with low risk, and \textbf{?} signifies an unknown status due to proprietary components. 

For each data path type, our security analysis is structured into three parts: (1) \emph{Methodology}, describing our experiments used to uncover the protection mechanism; (2) \emph{Observation}, explaining how the mechanism operates; and (3) \emph{Security Insight}, highlighting the security implications and mitigations. Due to space constraints, the full analysis is provided in Appendix~\S\ref{sec:datapath}. Below, we briefly summarize the key security findings for each data path.

\emph{1. CPU-\ac{GSP} \acf{RPC}.}
The NVIDIA kernel-mode driver communicates with the \ac{GSP-RM} through a physical \ac{RMAPI} \ac{RPC} interface. Because this communication occurs via shared system memory, which is visible and malleable to a compromised host, the commands and responses must be encrypted. To assess the security of this channel, we intercepted all \ac{RMAPI} \ac{RPC} transactions, decoded 1,588 command types using encodings extracted from the NVIDIA \ac{SDK}, and examined the shared staging buffer by dumping its physical memory. Our analysis reveals that only the command and status payloads are encrypted, while most \ac{RPC} metadata items in the staging buffers remain in plaintext and are exposed to potential attackers. This metadata leakage enables attackers to track \ac{RMAPI} invocation status, violating confidentiality guarantees. Although payload encryption with a session key protects \ac{RPC} command integrity, attackers can still modify metadata fields such as \code{readPtr} and \code{writePtr}, altering the order, repetition, or omission of \ac{RPC} calls and thus may partially undermine execution integrity. We recommend encrypting both \ac{RPC} payloads and metadata of command and status queues to provide stronger protection.

\emph{2. CPU-\ac{GSP} Memory Transfers.}
In the \ac{GPU-CC} mode, BAR2 access to the GPU's \ac{CPR} is blocked, so the NVIDIA kernel-mode driver uses CPU-\ac{GSP} \ac{DMA} to transfer memory. Because \ac{DMA} cannot access \ac{CVM}'s private memory directly, transfers must pass through an unprotected staging buffer in the shared memory. Memory read operations are initiated when the driver issues a \code{MEMMGR\_MEMORY\_TRANSFER\_WITH\_GSP} \ac{RPC} request, which prompts the \ac{GSP} to encrypt source data using the \code{gsp\_cpu\_dma} key. Following encryption, the data is transferred via \ac{DMA} into a staging buffer for driver-side retrieval and decryption. Conversely, for write operations, the driver first encrypts the data using the \code{cpu\_gsp\_dma} key and places it in the staging buffer. The driver then invokes the same \ac{RPC} to signal the \ac{GSP}, which performs a \ac{DMA} copy and decrypts the data into the \ac{CPR}. In our analysis, we intercepted all CPU-\ac{GSP} transfers across a \ac{CVM} lifecycle (4,394 transfers in total: 453 reads and 3,941 writes), logged execution times for transfer sizes from 8 to 4,096 bytes, and correlated latency with size to search for timing channels. The timing distribution is bimodal: for small transfers (8 to 256 bytes), \ac{RPC} overhead dominates and size has little effect. By contrast, for large transfers (4,096 bytes), execution times increase substantially, producing a size-dependent timing signal observable via the \ac{RPC} channel. This creates a potential timing side channel that could leak information about transfer size or activity. We recommend implementing constant-time \ac{RPC} handling or adding statistical noise for CPU-\ac{GSP} memory transfers to obfuscate these timing patterns.

\emph{3. GPU Memory Faults.}
Memory faults occur when a program accesses out-of-bound memory, dereferences null pointers, or writes to freed memory. In \ac{GPU-CC} mode, the hardware buffer for logging these faults resides inside the GPU's \ac{CPR}, making it inaccessible to the CPU. To examine the fault-handling path under \ac{GPU-CC}, we intentionally triggered faults in a customized CUDA program and instrumented both the NVIDIA kernel-mode and \ac{UVM} drivers to observe how fault packets are transferred and protected. We found that the \ac{UVM} driver allocates two shadow buffers in unprotected memory: one for replayable faults and one for non-replayable faults. The \ac{GSP} encrypts fault packets using \code{gsp\_cpu\_replayable\_fault} or \code{gsp\_cpu\_non\_replayable\_fault} keys and copies them into the appropriate shadow buffer, then signals the CPU via an \ac{RPC} event so the \ac{UVM} driver can verify and decrypt the packet. Unlike non-GPU-CC mode, both fault types now follow a unified workflow since the \ac{GSP} controls the hardware buffers. The only observed leakage is that the shadow buffer's \code{PUT} pointer is exposed to the CPU via BAR0 and is updated whenever a new fault is logged. However, since faults are infrequent, the risk of information leakage remains low.

\emph{4. \acf{UVM}.}
\ac{UVM} provides a unified CPU-GPU address space. But in \ac{GPU-CC} mode, where the GPU cannot access \ac{CVM}'s private memory and BAR2 access to the \ac{CPR} is blocked, \ac{UVM} must adopt a new secure design. \ac{GPU-CC} introduces a two-phase, multi-channel workflow involving \ac{SEC2} Channel, \ac{WLC}, and \ac{LCIC}. Pushbuffers and command queue structures are now relocated into the \ac{CPR}. To examine this new pipeline, we instrumented the NVIDIA kernel-mode and \ac{UVM} drivers and reconstructed the control flow. We observed that the \ac{SEC2} Channel initializes 16 \ac{WLC}/\ac{LCIC} pairs by verifying the signed static methods, with their schedules configured through \ac{SEC2} pushes. \ac{UVM} \ac{CE} pushes are then launched indirectly via \acp{WLC}: their \emph{decrypt-then-run} workflow transfers encrypted pushbuffers into the \ac{CPR} and decrypts them for execution, while encrypted tracking semaphores in staging memory allow the driver to monitor progress. Our security analysis shows that \ac{SEC2}'s command queue structures, specifically \code{GPPUT}, \code{GPFIFO}, and tracking semaphores, remain unprotected in shared memory, which could potentially allow attackers to redirect execution toward malicious pushbuffers. However, the \ac{SEC2}'s method-signing mechanism may still prevent modified GPU methods from being executed. Meanwhile, all critical \ac{WLC}/\ac{LCIC}/\ac{CE} data structures, including their mutable pushbuffers, command queues, and tracking semaphores, remain encrypted in staging buffers or confined to the \ac{CPR}.

\emph{5. Memory Scrubbing.}
Memory scrubbing ensures sensitive data in GPU memory is securely erased before GPU reuse. In \ac{GPU-CC} mode, a dedicated scrubber channel associated with the \ac{SEC2} engine is introduced to prevent tampering with scrub operations. To study this workflow, we intercepted \ac{SEC2} signing functions and backtracked scrub triggers within the memory manager. We found that when memory pages are freed, the driver creates a pushbuffer with a series of memset commands, signs them using the \code{cpu\_sec2\_hmac\_scrubber} key, appends authentication tags, and dispatches the signed pushbuffer to \ac{SEC2} engine via the scrubber channel. The \ac{SEC2} engine validates the signature before executing the scrub task. Our security analysis reveals two limitations: (1) scrub pushbuffers are signed but not encrypted, exposing their GPU methods to attackers on the host and (2) the tracking semaphores used by the \ac{SEC2} engine to signal task completion remain unprotected. These gaps allow attackers to observe scrub activity or tamper with completion signals, suggesting the need for stronger protection in the scrubber channel.

\emph{6. CUDA.} CUDA is NVIDIA's proprietary parallel computing platform, designed to offload compute-intensive portions of an application from the sequential CPU to parallel GPU cores. \ac{GPU-CC} must protect five categories of data involved in CUDA operations: (1) User data (e.g., datasets and model weights), (2) CUDA kernel code, (3) CUDA kernel arguments, (4) \ac{QMD} launch configuration structure, and (5) command queue structures used to transfer data and orchestrate kernel execution. To understand how \ac{GPU-CC} secures CUDA's data transfers, we instrumented the NVIDIA kernel-mode driver, the \ac{UVM} driver, and the \spec{OpenSSL} library to capture their interactions with CUDA programs. Our experiments reveal that the CUDA user-mode driver retrieves two sets of keys via the \code{GetKMB} API: \code{cpu\_sec2\_\{data,hmac\}\_user} keys, which are bound to the \ac{SEC2} engine, and \code{lce\{x\}\_\{h2d,d2h\}\_user} keys, associated with \acp{CE}. Analysis indicates that critical tasks, such as kernel execution and data movement, are primarily managed within the CUDA runtime and user-mode driver, with no visibility from the kernel space. Given that both the CUDA runtime and user-mode driver are closed-source, we make the following reasoned speculation based on the data that require special protection under \ac{GPU-CC}:

\speculation{With \ac{GPU-CC} enabled, all CUDA-related data transmission must be protected when passing over the untrusted PCIe interface. (1) Large data transmission, such as user data and CUDA kernel code, must be encrypted with \ac{CE}'s session keys, staged in shared buffers, and transferred via the \ac{CE} into the GPU's \ac{CPR} for decryption. Reverse transfers from GPU's \ac{CPR} to system memory should be similarly protected.  (2) Smaller but sensitive structures, including CUDA kernel arguments and \ac{QMD} launch configuration, can no longer be written in plaintext over PCIe. Instead, they should be encrypted by the CUDA user-mode driver and securely pulled by a GPU security engine (e.g., \ac{SEC2}), which verifies integrity and decrypts them within the \ac{CPR} prior to execution. (3) Command queue structures (e.g., \code{GPFIFO}, \code{GPPUT}, pushbuffers, and tracking semaphores) must either reside in the \ac{CPR} or remain encrypted in staging buffers, with secure work submission enabled through authenticated and encrypted channels managed by \ac{SEC2}.}

\nip{Key Takeaway.} \ac{GPU-CC} introduces stricter security requirements to protect data transmission. Yet much of the existing GPU software stack consists of legacy implementations that do not meet these expectations and must be updated individually. While recent changes secure most bulk command and data transfers, some sensitive metadata, timing behavior, and coordination signals remain exposed in the unprotected shared memory. These residual leaks can enable inference of computational behavior and in some cases even manipulation of operations, creating a partial loss of integrity. Extending protection to metadata and reducing observable timing differences would better strengthen the end-to-end pipeline security.
\section{Conclusion}
The NVIDIA \ac{GPU-CC} system aims to provide a secure execution environment for privacy-critical parallel computing workloads. However, its proprietary ecosystem, lack of public specifications, and design complexity hinder comprehensive security verification. This paper demystifies its inner workings to establish a foundation for future security research. We detail the underlying architectural mechanisms and present a series of in-depth experiments to assess security risks and identify potential attack surfaces.
\section*{Acknowledgments}
We are grateful to the anonymous reviewers for their feedback and suggestions. 
The authors at The Ohio State University were partially supported by NSF awards 2112471, 2207202, and 2348754. 
\bibliography{main}
\bibliographystyle{mlsys2026}
\appendix
\acresetall
\section{Security Analysis of Data Paths}
\label{sec:datapath}

\subsection{CPU-GSP RPC}
\label{ssec:rpc}
\nip{Methodology.}
The NVIDIA kernel-mode driver communicates with the \ac{GSP-RM} via the physical \ac{RMAPI} \ac{RPC} interface over a bi-directional channel. The driver sends commands to the \ac{GSP} for offloading specific tasks, while the \ac{GSP} responds with execution status and GPU events to notify the driver. Since the \ac{RPC} occurs over an untrusted medium that attackers on the host can examine and interpose, all data transmitted through the \ac{RPC} channels must be encrypted. 

We intercepted all physical \ac{RMAPI} invocations from the NVIDIA kernel-mode driver and captured their corresponding responses from the \ac{GSP}. Each command type follows a specific encoding, automatically generated by NVIDIA’s \spec{FINN} tool. To better understand the semantics of each \ac{RPC} invocation, we extracted the encodings of 1,588 \ac{RMAPI} commands from NVIDIA’s \ac{SDK} and replaced them in our captured log. 

\nip{Observation.}
The \ac{RPC} infrastructure is initialized during the \ac{GSP} engine construction phase. A shared memory region is allocated in unprotected system memory for communication, consisting of a \spec{physical address table}, a \spec{command queue}, and a \spec{status queue}. The \spec{physical address table}, which holds the physical addresses of all pages in the shared region, informs the GPU of the physical memory layout. The \spec{command queue} stores commands and their parameters sent from the driver to the \ac{GSP}, while the \spec{status queue} contains status information returned from the \ac{GSP} to the driver. The locations of these structures are communicated to the \ac{GSP} during initialization.

To send a command, the driver first sets up the command header with a \spec{sequence number} and \spec{element count}. The header also includes an \ac{AAD} buffer, an authentication tag, and a checksum, but it remains in plaintext. The driver then encrypts the command payload in private memory using the \code{cpu\_gsp\_locked\_rpc} key and computes a checksum over both the header and the encrypted payload, recording the value in the header. The header and encrypted payload are then copied into the \spec{command queue} in the staging buffer. The \ac{GSP} retrieves the command, verifies the checksum and the integrity of the encrypted payload, decrypts it using the same key, and executes the command.

To send a status update or notify the driver of an event, the \ac{GSP} encrypts the status using the \code{gsp\_cpu\_locked\_rpc} key and copies the status header along with the encrypted payload into the \spec{status queue} in the staging buffer. The driver retrieves the data from the \spec{status queue}, copies it to private memory, and verifies the checksum and the integrity of the encrypted payload. Then it decrypts the status payload and processes it accordingly.

\begin{figure}[!t]
\centering
\includegraphics[width=0.5\textwidth]{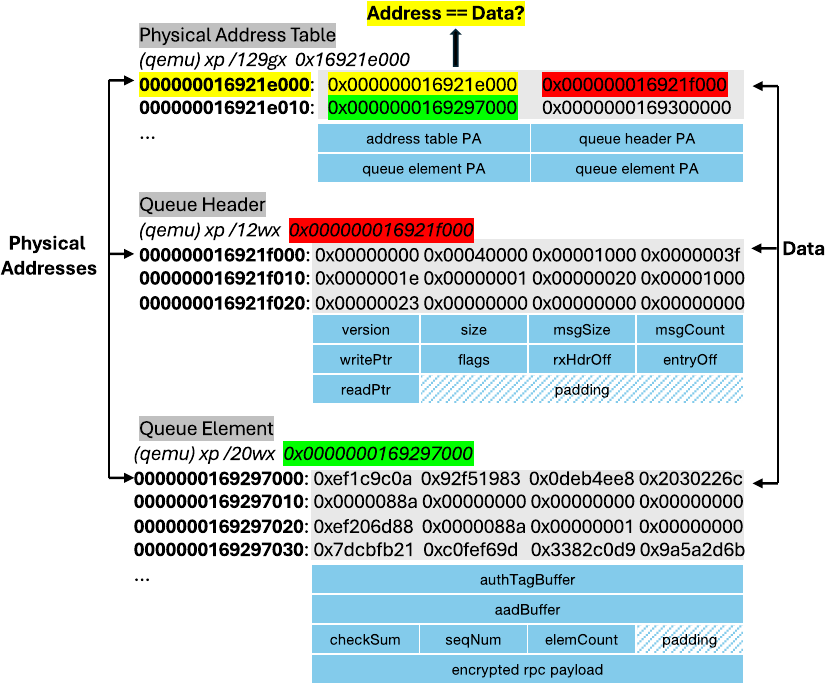}
\caption{Information leaks in \ac{RMAPI} \ac{RPC}: By traversing the memory addresses within the physical address table, which is allocated in unprotected system memory, adversaries can identify the plaintext metadata of queue and element headers.}
\label{fig:rpc}
\end{figure}

\begin{figure*}[!ht]
    \centering
    \begin{subfigure}{\textwidth}
        \centering
        \includegraphics[width=1\linewidth]{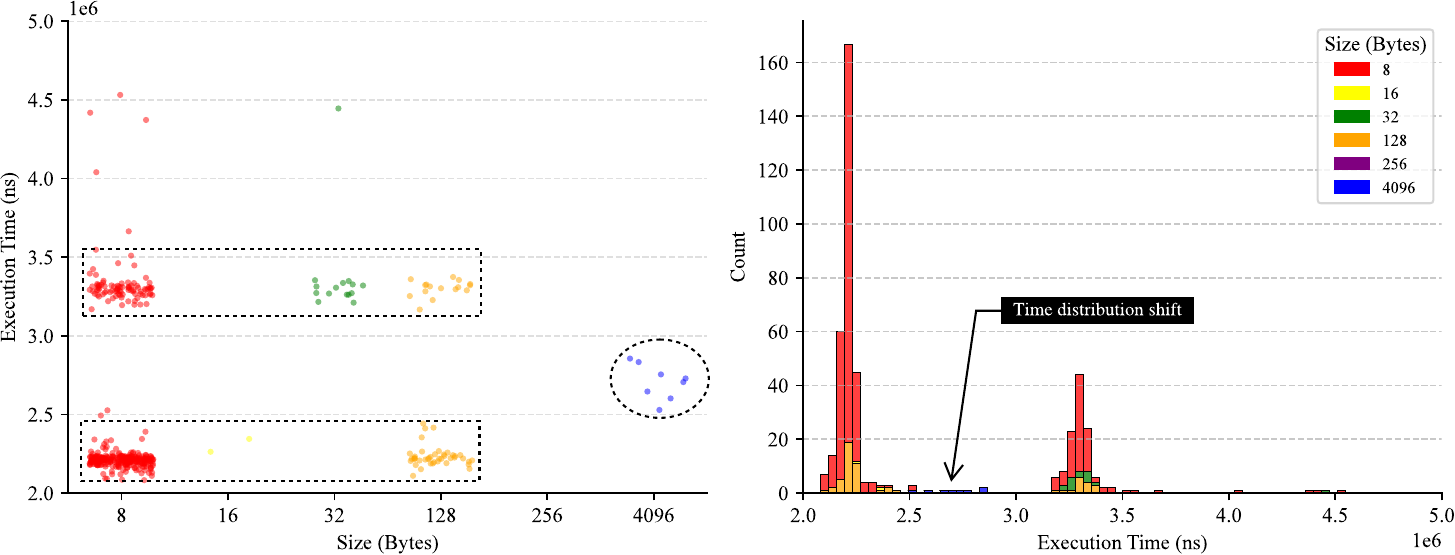}
        \caption{Read data from \ac{GSP}}
    \end{subfigure}
    \begin{subfigure}{\textwidth}
        \centering
        \includegraphics[width=1\linewidth]{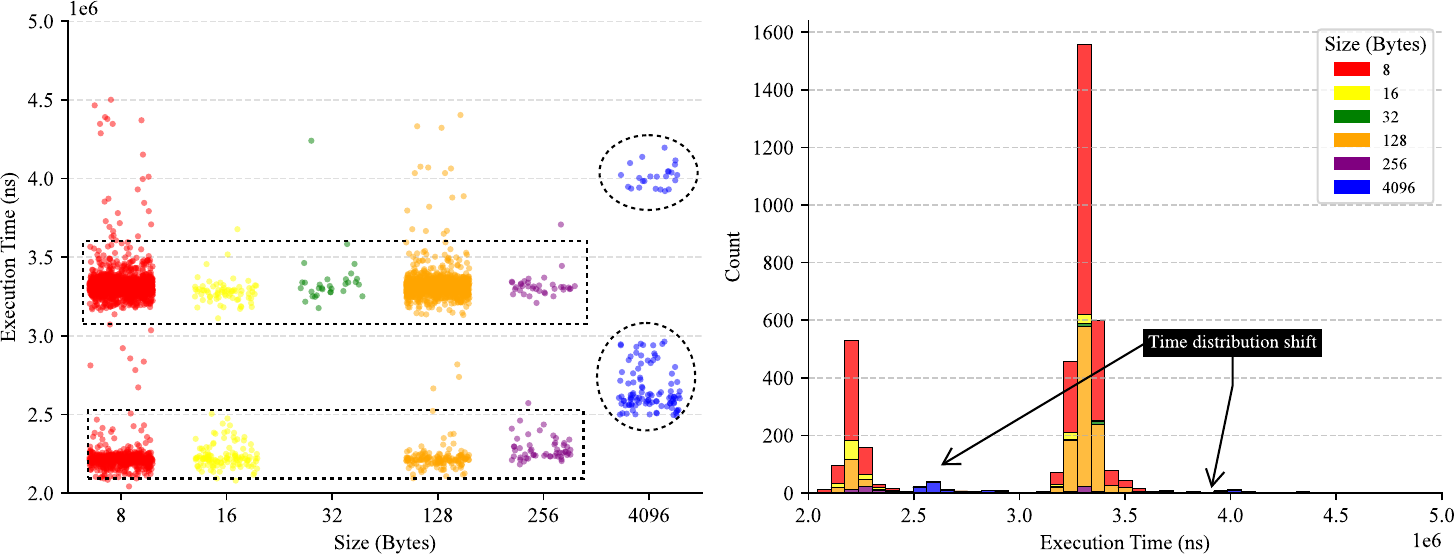}
        \caption{Write data to \ac{GSP}}
    \end{subfigure}
    \caption{Timing channels in CPU-\ac{GSP} memory transfers}
    \label{fig:time}
\end{figure*}
\nip{Security Insight.} It is important to note that only the command and status \ac{RPC} payloads are encrypted in the unprotected staging buffers, while the \spec{physical address table}, \spec{queue headers}, and \spec{queue element headers} remain in plaintext and are exposed to potential attackers on the host. Figure~\ref{fig:rpc} illustrates these data structures by dumping memory from their corresponding physical addresses in the \ac{KVM} hypervisor.

The physical address table consists of 129 entries (Figure~\ref{fig:rpc} shows the first four). The first entry stores the table’s own physical address, followed by one entry for the command queue’s header, 63 entries for command queue elements, one entry for the status queue’s header, and 63 entries for status queue elements. If adversaries gain access to the physical address table, they can easily locate all elements in the command/status queues. For example, the second entry (\code{0x000000016921f000}) points to the command/status queue's header, where the \code{readPtr} and \code{writePtr} fields indicate the next elements to read and write. These fields can then be used as indices in the address table to locate active elements in queues.

For instance, the third entry (\code{0x0000000169297000}) points to a specific queue element. Each element contains a header with fields such as \code{authTagBuffer}, \code{aadBuffer}, \code{checkSum}, \code{seqNum}, and \code{elemCount}, followed by the encrypted \ac{RPC} payload. The plaintext header can also reveal significant information about the currently active element. For example, if the \code{elemCount} has a value greater than one, it can be used to infer the \ac{RPC} type with high accuracy. 

A key challenge is how to locate the \spec{physical address table} within the large physical memory space of a \acf{CVM}. The vulnerability arises because the first entry in the address table stores its own physical address (e.g., \code{0x000000016921e000} in Figure~\ref{fig:rpc}). Given that the address table is page-aligned, adversaries can perform a memory scan in 4096-byte strides (the typical page size) across the entire physical memory. If a 64-bit physical address found during the scan \emph{equals} the value stored at that location, the attacker can uniquely identify the starting address of the physical address table, potentially exposing critical memory mappings.

Leakage of \ac{RPC} metadata enables adversaries to monitor \ac{RMAPI} invocation status, thereby may weaken confidentiality guarantees. While the \ac{RPC} payload is protected by the session key, which prevents direct compromise of its integrity, attackers can still manipulate the \code{readPtr} and \code{writePtr} fields in the metadata to alter the active elements. This can change the order, repetition, or omission of \ac{RPC} calls, thus may partially undermine execution integrity. To mitigate both leakage and manipulation, we recommend encrypting not only the \ac{RPC} payload but also the metadata of the command and status queues.

\subsection{CPU-GSP Memory Transfers}
\label{ssec:mt}
\nip{Methodology.} Once we enable the \ac{GPU-CC} mode, the BAR2 access to the GPU memory is blocked. Reads and writes to data within the \ac{CPR} should use alternative routes. For the memory transfers between \ac{CVM} and \ac{GSP}, the driver uses the CPU-GSP \ac{DMA}. As \ac{DMA} cannot access the private memory in the \ac{CVM} directly, the driver needs to allocate and map a staging buffer in the unprotected memory, which is not protected by the \ac{CVM}'s key. 

In our experiment, we intercepted all memory transfers between the \ac{CVM} and \ac{GSP} and recorded their execution time throughout the lifecycle of a \ac{CVM}. Our goal is to investigate how the source or destination aperture and the size of memory transfers impact execution time and whether the potential presence of timing channels can be exploited.

\nip{Observation.}
To read data from the GPU's \ac{CPR}, the driver sends a \code{MEMMGR\_MEMORY\_TRANSFER\_WITH\_GSP} \ac{RPC} request with parameters specifying the memory size, the source in the GPU's \ac{CPR}, and the destination in a staging buffer located in unprotected system memory. The protection for transmitting this \ac{RPC} request itself has been discussed in \S\ref{ssec:rpc}. Upon receiving the request, the \ac{GSP} encrypts the source data in the \ac{CPR} using the \code{gsp\_cpu\_dma} key and copies it via \ac{DMA} into the unprotected staging buffer. The driver then transfers the data to private memory and decrypts it using the same key.

To write data to the GPU's \ac{CPR}, the driver first encrypts the data in the \ac{CVM}’s private memory using the \code{cpu\_gsp\_dma} key and copies it to the unprotected staging buffer. The driver then sends the memory transfer request via the same \ac{RPC}, specifying the memory size, the source in the staging buffer, and the destination in the \ac{CPR}. Upon receiving the command, the \ac{GSP} copies the data via \ac{DMA} from the unprotected staging buffer, decrypts it using the same key, and writes it to the target address in the \ac{CPR}.

\nip{Security Insight.}
We recorded the execution time for 4,394 memory transfers between the CPU and \ac{GSP} throughout an entire \ac{CVM} lifecycle, consisting of 453 read operations and 3,941 write operations. Our goal is to analyze whether the transfer size impacts execution time. 

In Figure~\ref{fig:time}, we first illustrate the relationship between memory transfer sizes and execution time for both read and write operations in the two plots on the left. We can observe that large memory transfers, specifically those of 4,096 bytes, demonstrate a distinct shift in execution time compared to smaller transfers ranging from 8 to 256 bytes.

To visualize this statistical shift more effectively, we represent the data as histograms in the right-side plots of Figure~\ref{fig:time}. In these histograms, the horizontal axis denotes execution time while the vertical axis indicates the count of memory transfers within specific intervals. Distinct colors are used to identify the varying transfer sizes. The histograms reveal that both read and write operations exhibit a bimodal distribution of execution times. For small transfer sizes (8 to 256 bytes), the execution time is largely dominated by the inherent cost of the \ac{RPC} itself, making the effect of memory size relatively insignificant. However, for larger transfers (4,096 bytes, represented by blue bars), a noticeable shift occurs, with execution times increasing significantly compared to smaller transfers.

To mitigate potential timing side-channel attacks, where adversaries monitoring the \ac{RPC} channel in untrusted memory could infer sensitive information from execution time variations, we recommend implementing constant-time execution for \ac{RPC} invocations or introducing statistical noise to obfuscate timing patterns.

\subsection{GPU Memory Faults}
\label{ssec:faults}
\nip{Methodology.}
In \ac{GPU-CC} mode, handling GPU memory faults presents unique challenges due to the stricter memory access control. Typically, memory faults occur when a program attempts to access out-of-bound memory, dereference null pointers, or write to freed memory. In the non-GPU-CC mode, both the CPU-side drivers and GPU have access to a shared hardware buffer, allowing the GPU to log fault packets directly into the buffer, which the driver can then retrieve and handle. However, in \ac{GPU-CC} mode, this hardware buffer is placed within the GPU's \ac{CPR}, rendering it inaccessible to the CPU. As a result, a new mechanism is required to securely transfer fault packets through an untrusted interface.

To analyze this memory fault workflow in \ac{GPU-CC} mode, we developed a CUDA program designed to deliberately trigger memory faults. By instrumenting the NVIDIA kernel-mode driver and \ac{UVM} driver, we intercepted \ac{RPC} invocations used for setting up the staging buffers to deliver the fault packets and monitored the use of cryptographic keys related to memory faults. 

\nip{Observation.}
To meet the memory access control requirements in \ac{GPU-CC}, the \ac{UVM} driver requests the NVIDIA kernel-mode driver to allocate two shadow buffers, one for replayable memory faults and another for non-replayable memory faults. These shadow buffers reside in unprotected staging buffers, enabling the \ac{GSP} to write directly to them. The NVIDIA kernel-mode driver registers these shadow buffers with the \ac{GSP} through the \ac{RMAPI} \ac{RPC}, informing the \ac{GSP} of their structures. The \ac{GSP} then encrypts the fault packets using the \code{gsp\_cpu\_replayable\_fault} or \code{gsp\_cpu\_non\_replayable\_fault} key before copying them into the shadow buffers. Once the fault packets are stored, the \ac{GSP} sends a \code{MMU\_FAULT\_QUEUED} event via the \ac{RPC} return path, notifying the driver. The \ac{UVM} driver monitors this event and schedules an interrupt service routine to retrieve the encrypted fault packets from the shadow buffers. Each fault packet includes an \emph{authentication tag} and a \emph{valid} field (used as \ac{AAD}). The \ac{UVM} driver uses the corresponding key to verify and decrypt incoming packets, then parses them to handle memory faults effectively while preserving data confidentiality and integrity.

\nip{Security Insight.}
According to the comment~\cite{faultcomment} in the \ac{UVM} driver's source code, replayable faults originate from the \acp{SM}, while non-replayable faults come from other engines, e.g., \acp{CE}. In non-GPU-CC mode, these two types of faults follow different handling paths. The \ac{UVM} driver exclusively manages replayable faults and owns the hardware buffer that stores them. In contrast, handling non-replayable faults requires interaction with the NVIDIA kernel-mode driver via a specific shadow buffer since only the kernel-mode driver can access the hardware buffer for non-replayable faults.

However, in \ac{GPU-CC} mode, where the \ac{GSP} owns the hardware buffer and the CPU lacks direct access, both replayable and non-replayable faults rely on shadow buffers to store encrypted fault packets. Therefore, their handling paths become similar in \ac{GPU-CC} mode. Additionally, the \ac{GSP-RM} needs to update the shadow buffer's \code{PUT} pointer whenever a new fault packet is placed. NVIDIA repurposes the access counter registers mapped in BAR0 to expose this pointer to the CPU. This means that adversaries can also observe this value on the host. Given that memory faults occur infrequently, the risk of exposing the \code{PUT} pointer remains low. 

\begin{figure*}[!t]
\centering
\includegraphics[width=0.9\textwidth]{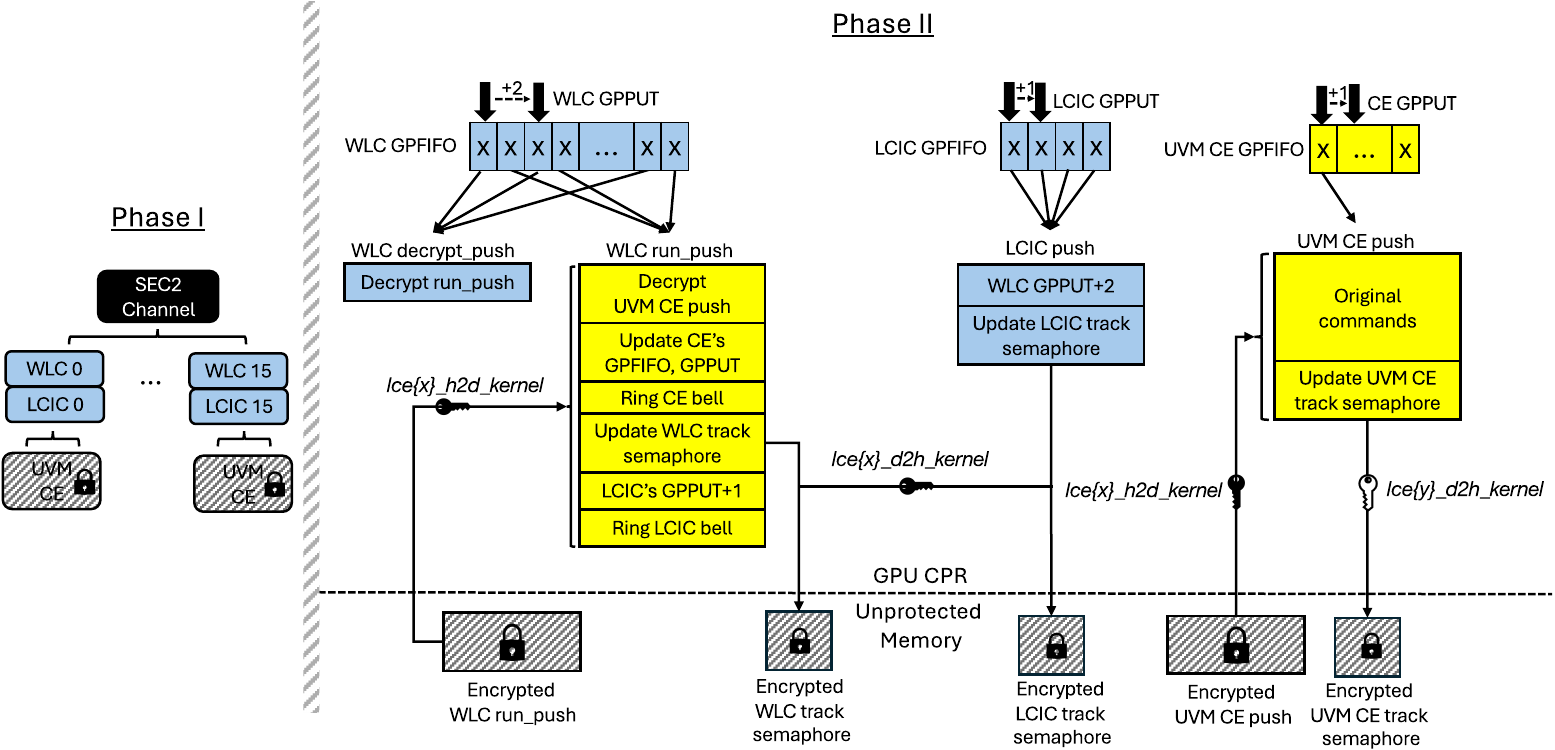}
\caption{The interactions among \ac{SEC2} Channel, \ac{WLC}, and \ac{LCIC}}
\label{fig:wlclcic}
\end{figure*}
\subsection{\acf{UVM}}
\label{ssec:uvm}
\nip{Methodology.}
NVIDIA's \ac{UVM} system provides a unified address space across system memory and GPU memory, enabling seamless memory access by managing memory migrations, page faults, and coherence automatically.

In the non-GPU-CC mode, the \ac{UVM} driver communicates with \acp{CE} through \ac{UVM} channels to conduct memory operations such as page copies or \ac{TLB} invalidations. Each \ac{UVM} channel owns specific regions of memory called pushbuffers, which are allocated in system memory and mapped for both CPU and GPU access. The driver writes sequences of GPU methods into these pushbuffers to prepare workloads for execution. A \code{GPFIFO} operates as a circular ring buffer that contains pointers to these pushbuffers. When the \ac{UVM} driver submits work, it populates a pushbuffer with methods and adds a pointer to the \code{GPFIFO}. The driver then increments the \code{GPPUT} pointer and triggers the doorbell to notify the \ac{CE} of the pending task. Subsequently, the \ac{CE} fetches and executes the methods from the \code{GPFIFO} before updating the \code{GPGET} pointer to reflect the current status. To synchronize these asynchronous operations, each \ac{UVM} channel has a tracking semaphore. The last method in every pushbuffer is a command to increment this semaphore, which the \ac{UVM} driver polls to confirm the successful completion of all GPU methods.

In the \ac{GPU-CC} mode, the GPU cannot access the private memory of the \ac{CVM} and BAR2 access to the GPU's \ac{CPR} is also disabled. Consequently, key components essential for \ac{UVM} operations, such as the pushbuffers, \code{GPFIFO} ring buffers, \code{GPPUT} pointers, and tracking semaphores, can no longer be directly accessed or modified by either the CPU or GPU, as shared memory is untrusted. To mitigate the risk of compromise or leakage, these components require additional security measures while residing in unprotected staging buffer.

The implementation of \ac{UVM} in \ac{GPU-CC} is significantly more complex, involving multiple phases and recursive interactions. To better understand its execution, we instrumented both the NVIDIA kernel-mode driver and the \ac{UVM} driver, restructuring the control flow into a clear temporal sequence. Our primary focus is analyzing the interactions between different engines, such as \ac{SEC2} and \acp{CE}, and examining how their derived keys protect specific target components within this new memory management model.

\nip{Observation.}
Three specialized channels have been introduced in \ac{UVM} for \ac{GPU-CC}: the \ac{SEC2} Channel, the \acf{WLC}, and the \acf{LCIC}. 

\emph{1. \ac{SEC2} Channel}: The \ac{SEC2} Channel is the first to be created, serving as the foundation for bootstrapping secure workload submission. Its primary role is to verify and set up the \ac{WLC} and \ac{LCIC}. There is only one \ac{SEC2} Channel, which is associated with the \ac{SEC2} engine and uses two cryptographic keys: \code{cpu\_sec2\_hmac\_kernel} and \code{cpu\_sec2\_data\_kernel}, which are used for signing methods and encrypting payloads respectively.  

\emph{2. \acf{WLC}}: \acp{WLC} are responsible for indirectly launching \ac{UVM} workloads. They process encrypted \ac{UVM} \ac{CE} pushes submitted by the \ac{UVM} driver, verify their integrity, and dispatch them to selected \acp{CE} for execution. The \ac{SEC2} Channel initializes 16 \acp{WLC}, corresponding to the maximum number of concurrent pushes. Each \ac{WLC} is associated with a logical \ac{CE} with two keys: \code{lce\{x\}\_h2d\_kernel} for host-to-device encryption and \code{lce\{x\}\_d2h\_kernel} for device-to-host encryption, where \code{\{x\}} denotes the index of the associated logical \ac{CE}. 

\emph{3. \acf{LCIC}}: Similar to \acp{WLC}, \acp{LCIC} are also initialized by the \ac{SEC2} Channel. Each \ac{LCIC} is paired with a corresponding \ac{WLC}, sharing the same logical \ac{CE} association, resulting in a total of 16 \acp{LCIC}. The \ac{LCIC} is designed to track the execution progress of its paired \ac{WLC}. 

The workflow can be divided into two phases: 

\emph{\underline{Phase I}: Setup of \ac{WLC}/\ac{LCIC} via \ac{SEC2} Channel.} The \ac{SEC2} Channel acts as the trust anchor for bootstrapping \acp{WLC} and \acp{LCIC}. \ac{SEC2}'s \code{GPPUT} pointer, \code{GPFIFO} ring buffer, and pushbuffers are allocated in an unprotected staging buffer and remain unencrypted. However, the methods in the \ac{SEC2} pushbuffers are signed with the \code{cpu\_sec2\_hmac\_kernel} key. Thus, the \ac{SEC2} engine can verify the integrity of the methods to detect any tampering.

With \ac{GPU-CC} enabled, the \code{GPPUT}, \code{GPFIFO}, and pushbuffers of \acp{WLC}/\acp{LCIC} must be placed in GPU's \ac{CPR} to protect them from unauthorized access. The \ac{UVM} driver populates an \ac{SEC2} push to set up these data structures for \acp{WLC}/\acp{LCIC}. This process repeats for all 16 \acp{WLC} and \acp{LCIC}. At this stage, the \acp{WLC}/\acp{LCIC} are launched but not yet operational. The next step is to set up their schedules. 

The \ac{WLC}'s schedule consists of two alternating pushes: (1) \code{decrypt\_push}, responsible for decrypting the next \code{run\_push} and placing it into \ac{CPR}. The \code{decrypt\_push} is static and remains unchanged. (2) \code{run\_push}, executing GPU methods to launch a \ac{UVM} \ac{CE} pushbuffer that contains the payloads for the \ac{UVM} operations, and advances the \ac{LCIC}'s \code{GPPUT} by one step for synchronization. The \code{run\_push} is mutable, as its parameters must be dynamically updated to accommodate different \ac{UVM} \ac{CE} pushes.

The \ac{LCIC} schedule is entirely static and simply advances the \ac{WLC}'s \code{GPPUT} by two steps ahead, ensuring that the \ac{WLC} always starts from the \code{decrypt\_push} in the next cycle.

Finally, the static parts (marked blue in Figure~\ref{fig:wlclcic}) of the \ac{WLC}/\ac{LCIC} schedules are uploaded and their \code{GPFIFO} entries are updated by the \ac{SEC2} push. Once this is complete, the \acp{WLC}/\acp{LCIC} are fully operational and ready for accepting \ac{UVM} \ac{CE} pushes.

\emph{\underline{Phase II}: Indirect Launch of \ac{UVM} \ac{CE} Pushes via \ac{WLC}.}
Once the \acp{WLC}/\acp{LCIC} are fully initialized, they can be selected at runtime to indirectly launch a \ac{UVM} \ac{CE} push. A \ac{UVM} \ac{CE} push's pushbuffer contains methods for executing a specific \ac{UVM} task, e.g., writing page table entries for an address range. 

To execute a \ac{UVM} \ac{CE} push, the \ac{UVM} driver selects a \ac{WLC} and encrypts the scheduled \ac{UVM} \ac{CE} push's pushbuffer using the \code{lce\{x\}\_h2d\_kernel} key of that \ac{WLC}. The encrypted pushbuffer is placed in the unprotected staging buffer. A corresponding \code{run\_push} for the \ac{WLC} is generated based on the parameters of the \ac{UVM} \ac{CE} push, encrypted, and also stored in the staging buffer. The \code{decrypt\_push}, pre-loaded during Phase I, executes first to decrypt the \code{run\_push} into \ac{CPR}. Once decrypted, the \code{run\_push} begins execution by decrypting the \ac{UVM} \ac{CE} push's pushbuffer, setting up the \code{GPFIFO} and \code{GPPUT} within \ac{CPR}, and triggering it to run on another logical \ac{CE}. The \ac{UVM} \ac{CE} push uses the \code{lce\{y\}\_h2d\_kernel} and \code{lce\{y\}\_d2h\_kernel} keys tied to the newly attached logical \ac{CE}, where \code{\{y\}} denotes the engine index. Finally, the \code{run\_push} signals the \ac{LCIC} to advance the \ac{WLC}'s \code{GPPUT}, preparing it for the next execution cycle.

Throughout this process, \ac{WLC}, \ac{LCIC}, and \ac{UVM} \ac{CE} update their respective tracking semaphores, which are then encrypted and stored in the unprotected staging buffer. The \ac{UVM} driver continuously monitors these encrypted semaphores to track execution progress. The decryption of tracking semaphores is handled by different keys: \code{lce\{x\}\_d2h\_kernel} for the \ac{WLC}/\ac{LCIC} pushes and \code{lce\{y\}\_d2h\_kernel} for the \ac{UVM} \ac{CE} push. 

\nip{Security Insight.}
The \ac{UVM} system relies on the signed push of the \ac{SEC2} Channel to bootstrap \acp{WLC}/\acp{LCIC}. The \ac{SEC2} push is not encrypted since it only consists of predefined methods without embedded secrets. However, the \code{GPFIFO}, \code{GPPUT}, and tracking semaphores of the \ac{SEC2} Channel reside in an unprotected staging buffer and are neither signed nor encrypted. Adversaries might exploit this by manipulating the \code{GPFIFO} and \code{GPPUT} to redirect the execution to a crafted pushbuffer. However, since the methods themselves are signed, the \ac{SEC2} engine can still detect and prevent tampered methods from being executed. Therefore, we consider such an attack to compromise confidentiality by exposing the fields in the \ac{SEC2} Channel and to partially undermine execution integrity. To mitigate this risk, we recommend encrypting these fields in \ac{SEC2} Channel as well.

The critical data structures for \ac{WLC}, \ac{LCIC}, and \ac{CE} command queues, such as \code{GPFIFO} and \code{GPPUT}, are securely placed in \ac{CPR}. The pushbuffers, including \ac{WLC}'s \code{run\_push} and \ac{UVM} \ac{CE} pushes, are encrypted while in the unprotected staging buffer before being transferred into \ac{CPR}. Unlike the \ac{SEC2} push, these buffers are mutable and contain sensitive information, including memory addresses and \ac{UVM} operations. Therefore, they require encryption when transmitted through an untrusted interface. Additionally, the semaphores used for signaling the \ac{UVM} driver are also encrypted while in unprotected memory before being passed to \ac{CVM}’s private memory, ensuring confidentiality and integrity for retrieving the execution status of command queues.

\subsection{Memory Scrubbing}
\label{ssec:scrub}
\nip{Methodology.}
Memory scrubbing is used for deliberate overwriting of memory contents to clear out sensitive information. This prevents the recovery of confidential data, which is particularly important in secure environments where old data could otherwise be exploited by unauthorized users. Typically, the NVIDIA kernel-mode driver establishes a scrubber channel with the \ac{CE}, where it submits GPU methods for memory scrubbing, allowing the \ac{CE} to execute the task. However, when \ac{GPU-CC} is enabled, a dedicated and secure channel is required to submit memory scrubbing methods to prevent tampering through the untrusted interface. Instead of using \ac{CE}, \ac{SEC2} is used for memory scrubbing when \ac{GPU-CC} is enabled, as it supports signed push.

In our experiment, we intercepted the signing function invocations in the \ac{SEC2} utility responsible for signing methods in the scrubber channel. We then backtracked the call trace to identify the origin of memory scrubbing requests within the GPU's memory manager. This approach provided a deeper understanding into the mechanism underlying GPU's memory scrubbing.

\nip{Observation.}
The NVIDIA kernel-mode driver establishes a scrubber channel linked to the \ac{SEC2} engine to handle memory scrubbing tasks securely. When the GPU's memory manager needs to free certain memory pages, it populates a pushbuffer with a sequence of methods for a secure memset operation using \ac{SEC2}. These methods are then submitted to the scrubber channel. 

To securely free large GPU memory allocations, the memory manager chops the massive request into smaller chunks, generating cryptographically signed methods for each. To ensure integrity, the methods in the pushbuffers are signed using the \code{cpu\_sec2\_hmac\_scrubber} key, and their HMAC digests are stored in the authentication tag buffers. The \ac{SEC2} engine verifies the signature and zeros out the physical memory, firing an internal semaphore after every single chunk to notify the driver that it can safely recycle that specific authentication tag's memory slot for the next command. On the very last chunk of the loop, a completion semaphore is appended, forcing the hardware to flush all physical caches and notify that the entire block has been destroyed.

\nip{Security Insight.}
The earlier design of the NVIDIA kernel-mode driver reused the \code{cpu\_sec2\_hmac\_kernel} key for signing methods during memory scrubbing operations. While in the latest driver versions, NVIDIA introduced two dedicated scrubber keys, i.e., \code{cpu\_sec2\_\{hmac,data\}\_scrubber}, for handling memory scrubbing tasks. It is worth noting that the \code{cpu\_sec2\_hmac\_scrubber} key is now used for signing methods, while the \code{cpu\_sec2\_data\_scrubber} key exists but remains unused. We make the following reasoned speculation about the key usage:

\speculation{Given that the pushbuffers and the authentication tag buffers in the scrubber channel are currently allocated in the staging buffers, which are not encrypted, the \code{cpu\_sec2\_data\_scrubber} key may eventually be used to protect these data structures in the future.}

Additionally, the \ac{SEC2} engine only supports decryption and does not provide encryption capabilities. Notably, while there are \code{cpu\_sec2\_*} keys for securing data sent to \ac{SEC2}, there are no corresponding \code{sec2\_cpu\_*} keys for protecting data from \ac{SEC2} back to the driver. This means that the semaphores used to notify the driver remain unprotected. The rationale behind \ac{SEC2}'s lack of encryption support is unclear and needs further clarification from NVIDIA. This could present a security risk, as attackers on the host could observe the unencrypted pushbuffers and authentication tag buffers (though they cannot tamper with commands due to integrity verification) and manipulate the unprotected semaphores.

\subsection{CUDA}
\label{ssec:cuda}
\nip{Methodology.}
Since the CUDA user-mode driver is not open-source, its execution flow in \ac{GPU-CC} remains opaque. To analyze its behavior, we intercepted its interactions with external components that we could instrument. Our approach involved developing a sample CUDA application that performed operations such as GPU memory allocation, host-to-device data transfers, and CUDA kernel launches. However, instead of using the CUDA runtime APIs, we directly invoked the low-level CUDA user-mode driver APIs to gain finer control over execution. For instance, we compiled the CUDA kernels into \ac{PTX} code and launched them via \code{cuLaunchKernel}.

To further dissect communication flows, we instrumented the NVIDIA kernel-mode driver and \ac{UVM} driver, capturing their interactions with the CUDA user-mode driver. Additionally, drawing insights from PipeLLM~\cite{tan2025pipellm}, which identified \spec{OpenSSL} APIs as being used for encryption, decryption, and MAC verification by CUDA, we also instrumented relevant \spec{OpenSSL} functions. By preloading a customized version of the \spec{OpenSSL} library, we intercepted all cryptographic events invoked by the CUDA user-mode driver, allowing us to observe how security mechanisms are enforced.

\nip{Observation.}
By intercepting the corresponding \spec{OpenSSL} functions, we are able to observe the key strings used by the CUDA user-mode driver and map them to the keys derived from the master secret negotiated during the \ac{SPDM} session. Our analysis reveals that the CUDA user-mode driver retrieves two categories of keys from the NVIDIA kernel-mode driver through the \code{GetKMB} API: (1) \code{cpu\_sec2\_\{data,hmac\}\_user} keys, associated with the \ac{SEC2} engine and likely used for encrypted data transfer and integrity verification. (2) \code{lce\{x\}\_\{h2d,d2h\}\_user} keys, tied to specific logical \acp{CE} and facilitate encrypted data transfers between CPU and GPU's \acp{CE}. It is worth noting that, since CUDA operates in user space, all these keys are post-fixed with \code{\_user}, indicating their privilege level. 

Based on our understanding of CUDA's execution lifecycle and CPU-GPU data movement patterns, we consider that the following five categories of data require specialized handling under \ac{GPU-CC}:

\emph{(1) User Data.} This refers to the actual data, e.g., training datasets and model weights, on which the CUDA kernel operates. The \code{cudaMemcpy*} runtime APIs manage data movement by invoking the GPU’s \ac{CE} to transfer data over the PCIe bus using \ac{DMA}.

\emph{(2) CUDA Kernel Code.} This consists of the compiled \ac{SASS} instructions executed by the GPU's \acp{SM}. During CUDA context initialization, the CUDA user-mode driver uses the \ac{CE} to copy the CUDA kernel code into GPU's device memory. At kernel launch, the memory address of the kernel code is embedded in the \ac{QMD}, enabling the hardware to locate and fetch the instructions.

\emph{(3) CUDA Kernel Arguments.} Kernel arguments are the parameters passed from the CPU to a CUDA kernel function. They include scalar values, dynamic dimensions, and pointers to user data. At launch time, these arguments are packed and copied into shared memory accessible by both the CPU and GPU. They are then transferred to Constant Bank 0, which is backed by a dedicated constant cache in the \ac{SM}.

\emph{(4) \acf{QMD}.} The \ac{QMD} is a bit-packed launch configuration structure that encodes grid and block dimensions, register and memory requirements, pointers to kernel code and arguments, and hardware state controls. It is copied into shared memory and read by the GPU's command processor to configure the \acp{SM} for kernel launching.

\emph{(5) Command Queue Structures.} The CUDA user-mode driver prepares GPU methods in pushbuffers and links them into corresponding \code{GPFIFO} command queues. These include queues targeting the \acp{CE} for transferring user data and kernel code, as well as queues targeting GPU's command processor to initiate kernel execution.

\nip{Security Insight.}
The CUDA user-mode driver has limited interactions with the NVIDIA kernel-mode driver, primarily occurring during: CUDA API/context initialization, kernel channel creation and key retrieval, and memory cleanup. However, the most critical operations related to data transmission and CUDA kernel launch remain self-contained within the proprietary CUDA runtime and user-mode driver, making them less observable. Given the limited information available, we make the following reasoned speculation regarding the types of data that should be protected during CUDA execution:

\speculation{\nip{User Data and CUDA Kernel Code.} User data and CUDA kernel code are typically transferred via the GPU's \ac{CE} using \ac{DMA} due to their large size. With \ac{GPU-CC} enabled, however, the \ac{CE} cannot directly access private memory within a \ac{CVM}. Instead, both user data and kernel code must first be encrypted using a session key (e.g., the \code{lce\{x\}\_h2d\_user} key) and placed in a staging buffer. The \ac{CE} may then transfer the encrypted data via \ac{DMA} into the GPU's \ac{CPR}, where it is decrypted. Conversely, data transferred from the \ac{CPR} back to system memory must be encrypted with the session key (e.g., \code{lce\{x\}\_d2h\_user} key) before being written to the staging buffer.

\nip{CUDA Kernel Arguments and \ac{QMD}.} Under \ac{GPU-CC}, the conventional approach, where the CPU directly writes plaintext kernel arguments and \ac{QMD} to the GPU via PCIe \ac{BAR}, is disallowed to defend against an untrusted hypervisor and PCIe bus. Instead, a pull-based model can be adopted in which the GPU retrieves cryptographically protected data. Specifically, the CUDA user-mode driver may encrypt the kernel arguments and \ac{QMD} using a pre-established session key (derived from \ac{SPDM} key negotiation) and place the ciphertext in the staging buffer. A GPU security engine (e.g., \ac{SEC2}) may initiate a memory transfer to fetch the encrypted data over PCIe, verify its integrity, and decrypt it securely within the GPU's \ac{CPR} just prior to kernel execution.

\nip{Command Queue Structures.} Command queue components, including \code{GPFIFO}, \code{GPPUT}, pushbuffers, and tracking semaphores, must either reside within the GPU's \ac{CPR} or remain encrypted when stored in staging buffers. A secure work submission mechanism, similar to that provided by \ac{SEC2} in the \ac{UVM} driver, may be employed. For instance, the \code{cpu\_sec2\_hmac\_user} key may be used to authenticate \ac{SEC2} push's methods, while the \code{cpu\_sec2\_data\_user} key may encrypt data sent to the \ac{SEC2} engine. Work submission may then proceed indirectly through \ac{SEC2} or other secure channels, such as \ac{WLC}, which itself is securely bootstrapped by \ac{SEC2}.}
\section{Future Directions}
\label{sec:future}
Here, we discuss the security features expected in upcoming \ac{GPU-CC} releases, along with the architectural and software changes required to support them.

\subsection{Multi-GPU Support}
As AI models continue to grow in size, a single GPU's memory often cannot accommodate the full model or dataset, necessitating Multi-GPU support for both inference and training tasks. On our NVIDIA H100 SXM5 platform, each H100 GPU is connected to all four NVSwitches via 4th-generation NVLinks, enabling high-bandwidth communication between GPUs.

In \ac{GPU-CC}'s threat model, adversaries capable of intercepting traffic at the NVSwitches or along the NVLinks between GPUs and NVSwitches may be able to observe or tamper with inter-GPU data transfers. While NVIDIA’s paper~\cite{dhanuskodi2023creating} discusses Multi-GPU support, the initial software releases only allowed a confidential GPU to be passed through to a single \ac{CVM}. Recent software releases have added support for Multi-GPU passthrough under the \spec{Protected PCIe} mode, which requires updates to the VBIOS, NVIDIA GPU driver, and CUDA\footnote{\spec{Protected PCIe} mode requires, at a minimum, VBIOS version \code{96.00.BC.00.01}, NVIDIA GPU driver version \code{570}, and CUDA version \code{12.8}.}. According to NVIDIA's OC3 presentation~\cite{nertney}, \spec{Protected PCIe} mode only supports encrypted data transfers over the PCIe interface. Inter-GPU data transfers over NVLinks and NVSwitches are not encrypted due to performance considerations. NVLink encryption is expected to be introduced in the \spec{Blackwell} architecture.

Supporting \ac{GPU-CC} in a Multi-GPU setting introduces new requirements for securely transferring data across GPUs. Specifically, peer-to-peer keys must be established to protect memory transfers between each pair of GPUs. A GPU cannot directly access another GPU's \ac{CPR}, as the interface between them is considered untrusted. Although \ac{CPR} occupies the majority of GPU memory, a small unprotected portion remains available and can serve as a staging buffer, which is accessible to other GPUs. To transfer data securely, the sending GPU must first encrypt the data and write it to the staging buffer. The receiving GPU can then copy the encrypted data from the buffer and decrypt it into its own \ac{CPR}.

\subsection{Trusted I/O Support}
One limitation of the current \ac{GPU-CC} architecture is that the GPU cannot directly access the private memory of a \ac{CVM}. Instead, data must be transferred through a staging buffer, introducing additional software-based encryption and decryption overhead. This not only complicates software development, particularly when determining which data objects to protect, but also increases the risk of leaving some data unprotected during transmission.

A promising direction to address this limitation is the development of Trusted I/O for confidential computing. Trusted I/O requires architectural support from both CPUs and GPUs. On the CPU side, technologies like Intel TDX Connect~\cite{tdxio} and AMD SEV-TIO~\cite{sevtio2023} aim to enable this capability. On the GPU side, NVIDIA's \spec{Blackwell} architecture is expected to introduce Trusted I/O support as well\footnote{Based on the release notes of NVIDIA kernel driver \code{590}, both \spec{Hopper} and \spec{Blackwell} currently only support encrypted data transfers through staging buffers, which is the mechanism described in \S\ref{sec:protection}. The initial support for PCIe link encryption and device authentication has recently been upstreamed to Linux kernel version \code{6.19} (February 2026). Additional Trusted I/O support is planned for the future kernel release cycles.}.

Intel and AMD's Trusted I/O technologies rely on a combination of standardized protocols and trusted components to ensure secure device integration and communication within confidential computing environments. Protocols such as \ac{TDISP}, \acf{SPDM}, and \ac{IDE} are central to this infrastructure. \ac{TDISP} governs the device interface lifecycle, while \ac{SPDM} is used for device authentication and establishing secure communication channels. \ac{IDE} provides encryption for PCIe traffic, securing both control and data paths. 

Trusted components include the \ac{TSM} and the \ac{DSM}. The \ac{TSM} is responsible for defining and enforcing security policies and managing secure communication with devices. The \ac{DSM}, a trusted logical component within the device, works in conjunction with the \ac{TSM} to establish secure channels. 

Device attestation is supported through \ac{SPDM} and the \ac{RIM}, which validate device identity and firmware integrity. When a device is assigned to a \ac{CVM}, the host initiates attestation by communicating with the device's \ac{DSM} using \ac{SPDM}. The \ac{TSM} verifies the identity and measurements of the device, checks them against security policies, and passes the evidence to the \ac{CVM} for final evaluation. If the device is verified, it is admitted into the trusted environment.

Once a device is trusted, it can perform \ac{DMA} to private guest memory. In Intel TDX Connect, this access requires the device to be included in the \ac{CVM}'s \ac{TCB}. The TDX module manages translation of device transactions to private memory, with the IOMMU enforcing these mappings and ensuring access is restricted to assigned devices. Secure PCIe transactions use IDE \acp{TLP} marked with a T-bit to signal trusted communication. In AMD SEV-TIO, \ac{DMA} access is permitted after the guest establishes trust in the device. The IOMMU maintains a \ac{SDT} that tracks the security attributes of devices and enforces access control accordingly.

Overall, both platforms ensure a secure and authenticated path for data and control between hosts, devices, and guest \acp{CVM}, building trust through layered verification and secure protocol enforcement.

\section{List of Acronyms}
\label{sec:acronym}
\ac{GPU-CC} uses a large number of acronyms. For clarity, we include a glossary of acronyms with cross-references to assist readers in navigating the paper.

\begin{acronym}[]\itemsep=0pt
\acro{AAD}{Additional Authenticated Data}
\acro{ABI}{Application Binary Interface}
\acro{ACM}{Authenticated Code Module}
\acro{ACPI}{Advanced Configuration and Power Interface}
\acro{AI}{Artificial Intelligence}
\acro{AES}{Advanced Encryption Standard}
\acro{AK}{Attestation Key}
\acro{ASID}{Address Space Identifier}
\acro{BAR}{Base Address Register}
\acro{BCB}{Bounds Check Bypass}
\acro{BIOS}{Basic Input/Output System}
\acro{BMC}{Baseboard Management Controller}
\acro{BROM}{Boot ROM}
\acro{CA}{Certificate Authority}
\acro{CC}{Confidential Computing}
\acro{CCA}{Confidential Compute Architecture}
\acro{CE}{Copy Engine}
\acro{CET}{Control Flow Enforcement Technology}
\acro{Ci}{Cryptographic Integrity}
\acro{CPU-CC}{CPU Confidential Computing}
\acro{CMRs}{Convertible Memory Regions}
\acro{CMR}{Convertible Memory Region}
\acro{CoT}{Chain of Trust}
\acro{CoVE}{Confidential VM Extension}
\acro{CPR}{Compute Protected Region}
\acro{CVM} {confidential virtual machine}
\acro{DCAP}{Data Center Attestation Primitives}
\acro{DIK}{Device Identity Key}
\acro{DHKE}{Diffie-Hellman Key Exchange}
\acro{DMA}{Direct Memory Access}
\acro{DRTM}{Dynamic Root of Trust for Measurement}
\acro{DoS}{Denial of Service}
\acro{DSM}{Device Security Manager}
\acro{ECC}{Error Correction Code}
\acro{EPC}{Enclave Page Cache}
\acro{EPCM}{Enclave Page Cache Map}
\acro{EPT}{Extended Page Table}
\acro{ESM}{Enter Secure Mode}
\acro{ES}{Encrypted State}
\acro{FSP}{Foundation Security Processor}
\acro{GE}{Graphics Engine}
\acro{GPA}{Guest Physical Address}
\acro{GPC}{Granule Protection Check}
\acro{GPT}{Granule Protection Table}
\acro{GPU}{Graphics Processing Unit}
\acro{GPU-CC}{GPU Confidential Computing}
\acro{GSP}{GPU System Processor}
\acro{GSP-BROM}{GSP BootROM}
\acro{GSP-FMC}{GSP First Mutable Code}
\acro{GSP-RM}{GSP Resource Manager}
\acro{GVA}{Guest Virtual Address}
\acro{HBM}{High‑Bandwidth Memory}
\acro{HE}{Homomorphic Encryption}
\acro{HKID}{Host Key Identifier}
\acro{HPA}{Host Physical Address}
\acro{IAS}{Intel Attestation Service}
\acro{IDE}{Integrity and Data Encryption}
\acro{ISR}{Interrupt Service Routine}
\acro{IOMMU}{Input/Output Memory Management Unit}
\acro{ISA}{Instruction Set Architecture}
\acro{IV}{Initialization Vector}
\acro{KET}{Key Encryption Table}
\acro{KVM}{Kernel-based Virtual Machine}
\acro{LCE}{logical copy engine}
\acro{LCIC}{Launch Confirmation Indicator Channel}
\acro{Li}{Logical Integrity}
\acro{LLM}{large language model}
\acro{LP}{Logical Processor}
\acro{LRU}{Least Recently Used}
\acro{LUKS}{Linux Unified Key Setup}
\acro{MAC}{Message Authentication Code}
\acro{MCTP}{Management Component Transport Protocol}
\acro{MEE}{Memory Encryption Engine}
\acro{ME}{Management Engine}
\acro{MIG}{Multi-Instance GPU}
\acro{MigTD}{Migration TD}
\acro{MKTME}{Multi-key Total Memory Encryption}
\acro{MMIO}{Memory-Mapped Input/Output}
\acro{MMU}{Memory Management Unit}
\acro{MRTD}{Measurement of Trust Domain}
\acro{MSK}{Migration Session Key}
\acro{MSR}{Model-Specific Register}
\acro{MTT}{Memory Tracking Table}
\acro{OTP}{One-Time-Programmable}
\acro{OVMF}{Open Virtual Machine Firmware}
\acro{OS}{operating system}
\acro{OCSP} {Online Certificate Status Protocol}
\acro{PAMT}{Physical Address Metadata Table}
\acro{PCCS}{Provisioning Certification Caching service}
\acro{PCE}{physical copy engine}
\acro{PCI}{Peripheral Component Interconnect}
\acro{PCK}{Provisioning Certification Key}
\acro{PCR}{Platform Configuration Register}
\acro{PCS}{Provisioning Certification Service}
\acro{PEF}{Protected Execution Facility}
\acro{PKI}{Public Key Infrastructure}
\acro{PSIRT}{Product Security Incident Response Team}
\acro{PSP}{Platform Security Processor}
\acro{PTE}{Page Table Entry}
\acro{PTX}{Parallel Thread Execution}
\acro{QE}{Quoting Enclave}
\acro{QMD}{Queue Metadata}
\acro{RATS}{Remote Attestation Procedures}
\acro{RDCL}{Rogue Data Cache Load}
\acro{RIM}{Reference Integrity Manifest}
\acro{RMAPI}{Resource Manager API}
\acro{RME}{Realm Management Extension}
\acro{RMP}{Reverse Mapping Table}
\acro{RoT}{Root of Trust}
\acro{RPC}{Remote Procedure Call}
\acro{RTMR}{Runtime Measurement Register}
\acro{S-IOV}{Scalable I/O virtualization}
\acro{SASS}{Streaming Assembler}
\acro{SDK}{Software Development Kit}
\acro{SDT}{Secure Device Table}
\acro{SE}{Secure Execution}
\acro{SEAM}{Secure-Arbitration Mode}
\acro{SEC2}{Secure Processor}
\acro{SEPT}{Secure EPT}
\acro{SEV}{Secure Encrypted Virtualization}
\acro{SGX}{Software Guard Extensions}
\acro{SLAT}{Second Level Address Translation}
\acro{SM}{Streaming Multiprocessor}
\acro{SMC}{Secure Multi-Party Computation}
\acro{SME}{Secure Memory Encryption}
\acro{SMM}{System Management Mode}
\acro{SNP}{Secure Nested Paging}
\acro{SoC}{System-on-Chip}
\acro{SPDM}{Security Protocol and Data Model}
\acro{SPR}{Sapphire Rapids}
\acro{SR-IOV}{Single Root I/O virtualization}
\acro{SVM}{Secure Virtual Machine}
\acro{SVN}{Security Version Number}
\acro{SVSM}{Secure VM Service Module}
\acro{TCB}{Trusted Computing Base}
\acro{TD}{Trust Domain}
\acro{TDCS}{Trust Domain Control Structure}
\acro{TDISP}{TEE Device Interface Security Protocol}
\acro{TDMR}{Trust Domain Memory Region}
\acro{TDR}{Trust Domain Root}
\acro{TDVPS}{Trust Domain Virtual Processor State}
\acro{TDX}{Trust Domain Extensions}
\acro{TEE}{Trusted Execution Environment}
\acro{TLB}{Translation Lookaside Buffer}
\acro{TLP}{Transaction Layer Packet}
\acro{TLS}{Transport Layer Security}
\acro{TME}{Total Memory Encryption}
\acro{TPM}{Trusted Platform Module}
\acro{TSM}{TEE Security Manager}
\acro{TVM}{TEE Virtual Machine}
\acro{TXT}{Trusted Execution Technology}
\acro{UD}{Undefined Instruction}
\acro{UEFI}{Unified Extensible Firmware Interface}
\acro{UVM}{Unified Virtual Memory}
\acro{VAPIC}{Virtual APIC}
\acro{VE}{Virtualization Exception}
\acro{VM}{virtual machine}
\acro{VMCS}{Virtual Machine Control Structure}
\acro{VMM}{Virtual Machine Monitor}
\acro{VMPL}{Virtual Machine Privilege Level}
\acro{VMX}{Virtual Machine Extensions}
\acro{VT}{Virtualization Technology}
\acro{vTPM}{virtual TPM}
\acro{WLC}{Work Launch Channel}
\end{acronym}
\end{document}